\documentstyle[12pt]{article}

\newcommand{\DATE}  {June, 1996}

\newcommand{\PPrtNo}
{
MSUHEP-60426 \\ CTEQ-604
}
\newcommand{\TITLE}
{ Improved Parton Distributions from Global Analysis of Recent Deep Inelastic
Scattering and Inclusive Jet Data }
\newcommand{\THANKS}
{
This work was partially supported by DOE and NSF
}
\newcommand{\AUTHORS}
{
H.~L.~Lai$^c$, J.~Huston$^c$, S.~Kuhlmann$^a$, F. Olness$^e$, J. Owens$^b$,
D. Soper$^d$, W.~K.~Tung$^c$, H. Weerts$^c$

}
\newcommand{\INST}
{
$^a$Argonne National Laboratory,
$^b$Florida State University,
$^c$Michigan State University,
$^d$University of Oregon,
$^e$Southern Methodist University
}
\newcommand{\ABSTRACT}
{
\begin{abstract}
The impact of recent precision measurements of DIS structure functions and
inclusive jet production at the Tevatron on the global QCD analysis of parton
distribution functions is studied in detail. Particular emphasis is placed on
exploring the range of variation of the gluon distribution $G(x,Q)$ allowed by
these new data. The strong coupling of $G(x,Q)$ with $\alpha_s$ is fully taken
into account. A new generation of CTEQ parton distributions, CTEQ4, is
presented. It consists of the three standard sets
(\mbox{\small {$\overline {\rm MS}$}}, {\protect \small DIS} and
leading order), a
series that gives a range of parton distributions with corresponding
$\alpha_s$'s, and a set with a low starting value of $Q$. Previously obtained
gluon distributions that are consistent with the high $E_t$ jet
cross-section are also discussed in the context of this new global analysis.
\end{abstract}
}

\renewcommand{\thefootnote}{\arabic{footnote}}

\begin{document}


\begin{titlepage}

\begin{tabular}{l}
\DATE
\end{tabular}
\hfill
\begin{tabular}{l}
\PPrtNo
\end{tabular}

\vspace{2cm}

\begin{center}
\renewcommand{\thefootnote}{\fnsymbol{footnote}}
{
\LARGE \TITLE  \footnote[2]{\THANKS}
}
\renewcommand{\thefootnote}{\arabic{footnote}}

\vspace{1.25cm}
{\large  \AUTHORS}

\vspace{1.25cm}

\INST
\end{center}

\vfill

\ABSTRACT                 

\vfill

\newpage
\end{titlepage}


\section{Introduction}

\label{sec:intro}

Lepton-lepton, lepton-hadron, and hadron-hadron interactions probe
different and complementary aspects of Quantum Chromodynamics. Each of
these interactions provides a window onto the elementary interactions
of quarks and gluons and probes the running coupling and quark masses.
The corresponding calculations are performed using perturbative Quantum
Chromodynamics (pQCD).  In addition these interactions probe the
partonic structure of hadrons, as represented by the parton
distribution and fragmentation functions \cite{handbook} \cite{OwTu}.
These functions are essentially non-perturbative.  There are, of
course, large areas of overlap between various processes, which provide
impressive consistency checks of the theory.
In the first approximation, lepton-lepton processes provide clean measurements
of basic parameters such as quark charges, the strong coupling $\alpha _s(Q)$%
, and fragmentation functions of partons into hadrons. Deep inelastic
scattering structure functions
and lepton-pair production cross-sections in
hadron collisions
provide the main source of information on the quark distributions
$f^q(x,Q)$ inside hadrons.
At leading order, the gluon distribution function $G(x,Q)$ enters directly in
hadron-hadron scattering processes with direct photon and jet final states.
In a global QCD analysis incorporating all these
processes, one tries to exploit the strengths of each process in a uniform
framework. Modern analyses are carried out to at least
next-to-leading order (NLO), thus \{$\alpha _s(Q),$ $f^q(x,Q),$ $G(x,Q)$\}
all contribute and mix in the theoretical formulas for each process.
However, the broad picture outlined above does reflect the main roles the
various processes play in the analysis.

Direct photon production has long been regarded as potentially
the most useful source of information on $G(x,Q)$. Fixed-target direct
photon data, especially those from WA70 \cite{wa70}, have been widely used
in existing global analyses. However, there are a number of theoretical
uncertainties which affect the predictions of the normalization and slope of
the measured direct photon $p_t$ spectrum. These effects include: (1) the
sensitivity of the theoretical calculations to the choice of factorization
and renormalization scales \cite{CtqDph}; (2) $k_t$ broadening
of the initial state partons due to soft gluon radiation \cite{CtqDph};
and (3) photon fragmentation uncertainties \cite{CtqDph}
and the related issue of photon isolation cuts \cite{FacIsoPh}.
When all these uncertainties are taken into account, existing direct
photon data do not place as tight constraints on the gluon distribution
as is commonly believed \cite{jet1}. Full exploitation of the potential
of this process in a QCD global analysis will require significant progress
in the understanding of the above issues.

An important process that is sensitive to gluons is jet production in
hadron-hadron collisions. In leading order, the cross-section is
proportional to $\alpha _s^2(Q)\,G(x,Q)\,G(x^{\prime },Q)$ and
$\alpha _s^2(Q)\,G(x,Q)\,q(x^{\prime },Q)$ for the gluon-gluon and
gluon-quark scattering subprocesses respectively. Experimental
measurement of various inclusive jet cross-sections has progressed to an
increasingly quantitative level in recent years. For instance, at the
Tevatron, good data on single jet production are now available over a wide
range of transverse energy, $15$ GeV $<E_t<450$ GeV \cite{CdfJets,D0Jets}.
NLO QCD calculations of jet cross-sections have also reached a mature stage
 \cite{EKS,AveGre,GieGlo}. Many issues relating to jet definition (which is
important for comparing theory with experiment) encountered in earlier
stages of jet analysis have been extensively studied and are better
understood. For the moderate to large $E_t$ range, the scale dependence of the
NLO inclusive jet cross section turns out to be relatively
small \cite{jetScleDep}. Thus, it is
natural that inclusive jet data should now be incorporated into a global QCD
analysis, and that these data should play a role in
constraining the gluon distribution $G(x,Q)$.

We have carried out a first systematic study of this problem using the
CTEQ global analysis framework \cite{ctqlng}.\footnote{%
A similar analysis was carried out earlier \cite{jet1}, focusing on the
interpretation of the ``high $p_t$ excess'' seen in the CDF jet measurement
 \cite{CdfJets}. See Sec.~\ref{sec:HiJet} for a discussion of the relation of
 \cite{jet1} to the current analysis.} In this paper, we shall focus on the
question: How well can the gluon distribution be determined as the result of
recent advances in experimental measurements? We discuss phenomenological
issues pertinent to extracting $G(x,Q)\,$ in the global analyses. These factors
are systematically taken into account in a series of analyses to gain
insight on the current range of uncertainties on $G(x,Q).$ We found that
recent, more precise, DIS data have a significant influence in narrowing
down the parton distribution functions (PDF's), including $G(x,Q);\,$and the
inclusion of inclusive jet data from hadron colliders further solidifies
knowledge on $G(x,Q)$ over a wide range of $x.$ As the result of this study,
we present new sets of CTEQ parton distributions (in $\overline{MS}$, DIS
and LO schemes) as well as a series of distributions which give a range of
variation of PDF's consistent with current data. We give quantitative
information on how these distributions compare to the data sets used in the
analysis. In addition, we provide a new set of PDF's with a low initial $%
Q_0^2=0.5$ GeV$^2;$ and we discuss the previously obtained gluon
distributions designed to accommodate the high $E_t$ jets \cite{jet1} in the
context of the CTEQ4 analysis.

\section{Issues on the determination of the gluon distribution}

\label{sec:issues}

In pQCD, the gluon distribution function is always accompanied by a factor
of the strong coupling (i.e. it appears as $\alpha _s\,G(x,Q)$), both in the
hard cross-sections and in the evolution equation for the parton
distributions. Thus, the determination of $\alpha _s$ and $G(x,Q)$ is in
general a strongly coupled problem. In principle, $\alpha _s$ can be
independently extracted from $e^{+}e^{-}$ collisions, or in sum rule
measurements in deep inelastic scattering.
$G(x,Q)$ can then be determined in a global analysis,
along with the quark distributions $f^i(x,Q)$,
by treating $\alpha _s$ as known.  Alternatively, one can try to determine
$\alpha_s$, $G(x,Q)$ and the quark distributions at once in a global analysis.
This relies on the
full ($x,Q$) dependence of the wide range of data to differentiate $\alpha
_s $ (which controls the overall $Q$-dependence of all quantities) from the
parton distributions (which depend on both $x$ and $Q$). This method is not
as ``clean'' as the first approach, and it will not become precise until the
global analysis system has become better constrained. Eventually, however, it
is important to demonstrate that the same value of $\alpha _s$ consistently
describes all the processes included in the global analysis. Hence, the two
approaches are indeed complementary.

It is well known that, at present, the value of $\alpha _s$ determined at
high energy colliders, especially LEP, is generally higher than that
obtained from analyses of fixed-target DIS data \cite{alphasRev}. Since
global QCD analyses are up to now dominated by the copious high statistics
DIS data, they favor values of $\alpha _s$ close to the lower ``DIS value''.
This situation may change when more and more quantitative results from
hadron collider processes, such as inclusive jet and direct photon
production, are included in the global analysis. In the following, we shall
explore the range of variation of $G(x,Q)$ when the value of $\alpha _s$ is
varied within the currently accepted region, which we shall take to be $%
0.105<\alpha _s(M_Z)<0.122.$ \cite{alphaRange} The problem of the
determination of $\alpha _s$ in global analysis and the question about
consistency of $\alpha _s$ among different processes will be considered in a
subsequent study \cite{CtqAls}.

For a quantitative study of $G(x,Q)$, another relevant consideration is: How
does the choice of parametrization of the initial gluon
distribution at some $Q=Q_0$ affect the results? All global analyses use a
generic form:
\begin{equation}
G(x,Q_0)\,=A_0\,x^{A_1}\,(1-x)^{A_2}\,P(x;A_3,..)  \label{iniGlu0}
\end{equation}
with $A_{1,2}$ being physically associated with small-$x$ Regge behavior and
large-$x$ valence counting rules respectively; and $P(x;A_3,...)$ being a
suitably chosen smooth function depending on one or more parameters. In
general, both the number of free parameters and the functional form can have
an influence on the global fit. In the CTEQ3 analysis \cite{ctqlng}, an
effort was made to minimize the number of free parameters, resulting in an
economical set whereby $A_1^G=A_1^{sea},$ and $P_{CTEQ3}(x;A_3)=1+A_3\,x.$
We shall refer to this choice as the {\em minimal set} in the following
discussions. In the literature, more degrees of freedom have been assigned
to $G(x,Q_0).$ For instance, in CTEQ2 \cite{ctqlng} and in recent MRS
fits \cite{mrsg},
$A_1^G$ is allowed to vary independently of $A_1^{sea};$ and
the function $P$ contains one more free parameter than $P_{CTEQ3}$: $%
P_{CTEQ2}(x;A_3,A_4)=1+A_3x^{A_4};\,P_{MRS}(x;A_3,A_4)=1+A_3\,\sqrt{x}+A_4x.$
Since two extra degrees of freedom are added, we shall refer to this
class of parametrization as {\em (m+2)} -- i.e. {\em %
minimal plus two}. The more general parametrization clearly allows a wider
range of variation of $G(x,Q_0).$ Some pertinent questions are: whether
these general parametrizations are required by current data; and do these
parametrizations give a good indication of the range of variation of $%
G(x,Q_0)?$ We shall investigate these questions in some detail
in the next two sections.

Finally, although the PDF's determined from global analysis should, in
principle, be universal, they could, in practice, depend on the choice of
data sets -- in particular, on the choice of ``$Q_{cut}$'' values that
specify the minimum hard physical scale ($Q,\,P_t,...)$ required for data
points in the various physical processes to be included in the fit. If the
NLO QCD theory is truly applicable in the kinematic range of the data, the
parton distributions should be insensitive to the value of $Q_{cut}.$ Since
current theory does not predict what value $Q_{cut}$ should take for each
process, this point has to be investigated phenomenologically.


\section{Impact of recent DIS data on the global analysis of parton
distributions}

\label{sec:nojet}

Since
the publication of the CTEQ3 analysis, more accurate and extensive DIS data
from NMC \cite{NewNmc} and HERA \cite{NewH1,NewZeus} as well as new data from
E665 \cite{E665} have become available. In comparing the new data with NLO
QCD $F_2$ computed from CTEQ3M distributions, we find general agreement,
except for the small-$x$ region where the more precise recent data show
deviations from the theory curves. This is shown in
Fig.~\ref{fig:NmcHacM}
for the NMC and H1 data sets respectively.%
\footnote{Results are similar for E665 and ZEUS.
Comparison with the full data sets will be presented later, cf.
Fig.~\ref{fig:DIScdM}
}
Thus, we first update the CTEQ3 analysis under several different conditions,
in order to study the impact of these new DIS data on the global analysis of
parton distributions, especially the extraction of the gluon distribution.

The magnitude of the uncertainty in $G(x,Q_0)$ due to the
current uncertainty on
$\alpha _s,$ will be investigated by systematically varying the value of $%
\alpha _s$ over the interval $0.105<\alpha _s<0.122$, as mentioned in the
previous section. We shall use the short-hand $\alpha _s$ for $\alpha _s(M_Z)
$ throughout. In terms of {\em QCD Lambda} values, this range of $\alpha _s$
corresponds to $100<\Lambda _5^{\overline{MS}}<280$ (MeV) and
$155<\Lambda _4^{\overline{MS}}<395$ (MeV).  We shall in
general use the $\overline{MS}$ scheme in NLO QCD.

To provide a base-line for comparison, we first obtain a series of such fits
under identical conditions and using the same data sets (i.e. pre-1995) as
in the CTEQ3 analysis \cite{ctqlng}. We shall refer to this as the A-series.%
\footnote{%
This series of fits were originally obtained in 1994. They have been used in
various phenomenological studies related to gluon distributions and $\alpha
_s$ determination conducted by CTEQ, CDF, and D0 Collaborations. They have
not been formally published.} By definition, the best fit in this series is
the published CTEQ3M fit with $\alpha _s=0.112$ $(\Lambda _5^{\overline{MS}%
}=158$ MeV). A comparison of the gluon distributions that correspond to
these values of $\alpha _s$ are presented in Fig.~\ref{fig:GluAa}.
In
order to render the differences in the various regions of $x$ visible over
the range $10^{-4}<x<1$, part (a) highlights the small-$x$ region by
plotting $x\,G(x,Q)$ against $\log \,x$, part (b) accentuates the medium-$x$
range by plotting $x^2\,G(x,Q)$ vs. $\log \,x$,\footnote{%
Note that, since $x^2G(x)\cdot d\log x=xG(x)\cdot dx=$ momentum fraction
carried within $dx$, each curve in this plot directly depicts the
distribution of the momentum fraction carried by the gluon for that set.}
and
part (c) emphasizes large-$x$ by plotting $x^2\,G(x,Q)$ vs. $x.$ For the
many detailed comparisons to follow, these separate plots, though
conventional, will prove to be rather cumbersome. We consolidate them into
one single less-conventional plot in Fig.~\ref{fig:GluA} in which all curves
are normalized by the function $x^{-1.5}(1-x)^3,$ which takes out most of the
singular (rapidly vanishing) factors at small (large) $x$.
The scale for the
abscissa is chosen to be a function of $x$ which smoothly interpolates
between $\log x$ (at small $x)$ and $x$ (at large $x)$ so that the behavior
of $G(x,Q)$ over the full $x$ range is more evenly displayed. We see that
all of the features seen in the three plots of Fig.~\ref{fig:GluAa}
are evident in
this single figure. This will be the format of choice in most subsequent
comparisons.

We see in Fig.~\ref{fig:GluA} that, in the region $x>0.05$ where the largest
concentration of data used for the fit lie, increasing values of $\alpha _s$
lead to decreasing values of $G(x,Q)$ -- as expected (particularly for the
direct photon data) since the product of the two enters into most
cross-section and evolution kernel formulas.\footnote{%
The order is reversed for small $x,$ because of the momentum sum rule.} As
noted before, in the CTEQ3 analysis, and therefore in this series of fits, the
initial gluon distribution function is parametrized {\em minimally} as
\begin{equation}
G(x,Q_0)=A_0\,x^{A_1}\,(1-x)^{A_2}\,(1+A_3x)  \label{IniGlu1}
\end{equation}
with $A_1$ set to be the same as that of the sea quarks. Hence there are 3
free gluon parameters -- $A_{0,2,3}$ -- in the fit.
For each $\alpha_s$, we found the best solution to be quite stable against
perturbations in the fitting procedure and starting parameters, indicating
the parametrization and the experimental constraints are well-matched.
This also results in an orderly variation of
$G(x,Q)$ as $\alpha _s$ is varied, as seen in the figure.
If one takes the range of $\alpha _s$ used here as
representing the current uncertainty on $\alpha _s$, then the spread of the
gluon distribution shown in Fig.~\ref{fig:GluA} gives the corresponding
uncertainty on $G(x,Q)$ (based on the data available prior to 1995, and
on the variation of $\alpha _s$ alone). We
should mention that, although quark distributions are allowed to vary
freely, the valence quark distributions remain practically the same for all
of the fits in this series, as they are very much pinned down by the
precision DIS data in the region where they dominate the structure
functions. On the other hand, the sea quark distributions couple to $G(x,Q);$
thus they do show a systematic variation with $\alpha _s,$ although the
variation is somewhat reduced compared to that of the gluon.

Next, we investigate the impact of the new DIS data from NMC \cite{NewNmc},
E665 \cite{E665} and HERA \cite{NewH1,NewZeus} on $F_2$ by repeating the
same study, with the new data sets replacing the original ones. The
resulting series of fits is called the B-series.
The quality of these fits (measured in $\chi ^2$ values)
are similar to those of the A-series. Six representative gluon
distributions in this series are shown in Fig.~\ref{fig:GluB} along with
that of CTEQ3M for reference.
It is rather striking to note that the spread
in $G(x,Q)$ observed above in the small-$x$ ($<0.01$) region has been
practically eliminated. This is precisely the region covered by the HERA
experiments. In addition, the new gluons are shifted down from those of the
A-series in the region $0.05<x<0.3$ where all three DIS experiments
contribute. At first glance, this may appear surprising in view of the
conventional wisdom that $F_2$ data are only sensitive to quarks, not
gluons. However, we must realize that, first, in the small-$x$ region $%
G(x,Q)\,$is quite large---typically about 20 times bigger than the quark
distributions---thus it has a strong influence, directly and indirectly, on
all physical quantities through the hard cross-section and the evolution
equation. Moreover, these fits use the minimal parametrization, including
the constraint $A_1^G=A_1^{sea}$ which strongly couples the behavior of $%
G(x,Q)\,$at small-$x$ to that of sea quarks. Thus, the much better
determined $G(x,Q)$ just reflects the improved accuracy of new data in this
region. We note also, the large-$x$ behavior of the new series is somewhat
different from the A-series, even if there are no new data in that region.
This must be due to the indirect effect of the required changes below $x=0.1,
$ induced by the restrictive functional form Eq.~\ref{IniGlu1}, and the
constraint imposed by the momentum sum rule. We should point out that the
absolute value of the gluon distribution in the region above $x=0.5$ is very
small (about $10^{-3}$ compared to its value at $x=0.1$); thus the
significance of the observed differences should not be over-emphasized.

The minimal parametrization for $G(x,Q_0)$ used above was originally
chosen in the CTEQ3 analysis for its economy -- all data sets included in
these global analyses can be reasonably well fitted with this form. This
does not prove that the true gluon distribution must fall within the range
shown above; in particular, the true $G(x,Q_0)$ may be more complicated than
can be represented by this parametrization. (For instance, all global
analyses find it necessary to use one more parameter to describe the valence
quarks.) Only experiments probing $G(x,Q)$ in a different way can tell
whether our results so far are adequate. Before turning to such additional
input, we can obtain a different estimate of the uncertainty on the gluon
distribution that is complementary to the width of the ``band'' shown in
Figs.$~$\ref{fig:GluA}-\ref{fig:GluB}. We adopt the more general
``(m+2)'' parametrization of $G(x,Q_0)$ already used in
CTEQ2:
\begin{equation}
G(x,Q_0)=A_0\,x^{A_1}\,(1-x)^{A_2}\,(1+A_3x^{A_4})  \label{IniGlu2}
\end{equation}
In addition to introducing the new parameter $A_4$ compared to Eq.~\ref
{IniGlu1}, the parameter $A_1$ is untied from $A_1^{sea}$ and treated
as free. This results in a new series of fits, called the C-series.

With two more free parameters than in the B-series, one would expect (i) to
fit the collective data ``better'' than before and (ii) to find
an increased range
of variation of the gluon distribution. Indeed, the $\chi ^2$ for the fits
decreased slightly (by about $10$ ($/1000~pts.$)) compared to the corresponding
ones in the B-series. The gluon distributions at $Q=5$ GeV in this series
for 6 values of $\alpha _s$ is shown in Fig.~\ref{fig:GluC}.
First, we see
that the range of variation of $G(x,Q)$ in this series is much wider as
compared to that of series B, although both include the same improved DIS
data. In
particular, in the small-$x$ region the very narrow range in series B is
very much opened up by the freeing of the $A_1$ parameter for the gluon --
since now $q(x,Q)$ and $\alpha _sG(x,Q)$ can vary independently, the
measured $F_2$ (which depends on both) no longer constrains each piece
tightly as in the B-series. Secondly, we note that the gluon distribution does
not vary in a systematic manner as the $\alpha _s$ value is
varied -- in contrast to the well-constrained case in series A and B.
Further study has indicated that, unlike in the other cases, small changes in
the fitting process can lead to different solutions for some values of
$\alpha_s$. This
suggests that the fits are not entirely stable; or, in other words, the
system becomes somewhat under-constrained with the two extra parameters
introduced.

These observations point to the need for more experimental input in order to
better measure the gluon distribution. We need new data to determine
whether the additional degrees of freedom associated with $A_1^G$ and $A_4^G$
are required for the true gluon or whether the restricted form used in
series B is already sufficient. If $A_1^G$ and $A_4^G$ are required, these new
data could help to stabilize the fits found in the C-series and hence shed
light on the possible range of $G(x,Q)\,$ allowed. From the discussion given
in the introduction, it is clear that inclusive jet production data could
be used to help resolve these issues, as we will show in the next section.
To conclude this section, Table~\ref{tbl:Fits} summarizes the above
described three series of global fits, as well as those including jet data
to be discussed next.

\section{Comparison with New Inclusive Jets Cross-section}

\label{sec:Jets}

For studying the impact of inclusive jet production cross-section, we use
the recent measurement of $d\sigma /dE_t$ from the CDF \cite{CdfJets} and D0
 \cite{D0Jets} Collaborations. The preliminary data obtained in run IB of the
Tevatron by the two experiments are shown in Fig.~\ref{fig:JetData}.
Although data are available for $15$ GeV$<E_t<450$ GeV, we will include in
our NLO QCD analysis only data ${\em above}${\em \ 50~}GeV because there are
a number of potential theoretical and experimental problems that may affect
the proper comparison between NLO QCD theory and data for lower $E_t$. These
include (1) scale uncertainty of NLO QCD calculations, which becomes
non-negligible at low $E_t$ (cf. Fig.~\ref{fig:JetUncer}a);
(2) ambiguities in the definition of the ``underlying event''
coming from the proton-antiproton
remnants (cf. Fig.~\ref{fig:JetUncer}b); (3) possible problems in the
match between theoretical and experimental jet definitions,
such as fragmentation
products outside the jet cone; (4) $k_t$ broadening of the initial state
partons \cite{CtqDph}; and (5) non-perturbative corrections to the theory,
which could be of order $1/E_t$ rather than $1/E_t^2$ \cite{NonPer}.
All of these affect
low $E_t$ jets much more than high $E_t$ jets, as will be illustrated by two
examples, one theoretical and one experimental.
Fig.~\ref{fig:JetUncer}a shows
the scale-dependence of the NLO QCD calculation as a function of $E_t$: the
theoretical inclusive jet cross-section is shown for several choices of the
renormalization and factorization scale ($\mu =\mu _R=\mu _F)$ normalized to
our standard choice $\mu =$ $E_t/2.$\footnote{%
The theoretical calculations of jet cross-section in this paper are carried
out using the EKS program \cite{EKS}.}  
For low $E_t,$ the ratio becomes
large and unstable; above $50-75$ GeV, the different choices are within 10\%
and stay relatively constant---they amount to shifts in the overall
normalization of the cross-section. 
Fig.~\ref{fig:JetUncer}b shows the
percentage effect on the inclusive jet cross-section
in the CDF experiment due to a $\pm 30$ \% change
in the underlying event correction (in Run IA). 
Again, the uncertainty becomes
large below $50-75$ GeV.

To emphasize the quantitative aspects of the subsequent analysis, the measured
steeply falling $d\sigma /dE_t$ is normalized to the NLO QCD theoretical
expectation using the CTEQ3M parton distributions (solid horizontal line)
and displayed in Fig.~\ref{fig:JetcM} on a linear plot (with statistical
errors only on the data points).
In Fig.~\ref{fig:JetcM}, we have taken into account the slightly
different pseudo-rapidity coverage of the two experiments ($0.1<|\eta |<0.7$
for CDF vs. $|\eta |<0.5$ for D0) by normalizing each data set with respect
to the theory values computed with the corresponding $\eta $ range. In
addition, we have allowed a small overall normalization of the two data
sets, well-within the quoted uncertainties, for this comparison.
This figure shows that the two data sets agree quite well over the
entire $E_t$ range, especially when considering the quoted
systematic uncertainties (not shown).
See Ref.~\cite{jetnote} for more discussions.
We will discuss the experimental systematic uncertainties in the proper
context of the ``range'' of gluon distributions later in this
paper.  The rise of the data points at
high $E_t$ values over the CTEQ3M expectation, more noticeable for the CDF
points, has been the subject of much recent discussion and speculation \cite
{CdfJets,jet1,HiJetSp}. We will comment on this issue in the context of the
global analysis conducted in this paper in a later section.

Since most inclusive jet data are collected in the central rapidity region,
the $x$-value of the PDF's probed is around $x_t=2E_t/\sqrt{s}.$ For $50$
GeV $<E_t<450$ GeV, the $x$ range is approximately $0.06-0.5.$ Over this
range, the relative importance of the three parton subprocesses --
quark-quark, quark-gluon, and gluon-gluon -- shifts continuously from being
gluon-dominated to quark-dominated, as illustrated in Fig.\thinspace \ref
{fig:SubProc}. We should also keep in mind that these jet data probe hadron
structure at much higher momentum scales than fixed-target experiments. Due
to the nature of the QCD evolution equation, parton distributions at these
high momentum scales are determined by those at lower scales and higher $x$
values. Thus the effective $x$-range in $G(x,Q_0)\,$ for some $Q_0$, say $1.6
$ GeV used in our analysis, probed by these jet data extends to much higher
values than the nominal values mentioned above. Since the quark
distributions throughout this range are very well pinned down by DIS
experiments, one expects the jet data to be particularly useful in
constraining the gluon distribution. The value of $\alpha _s$ has
considerable influence on the gluon determination for several reasons.
First, the cross-section for medium $x_t$ is proportional to $\alpha
_s^2\,G^n(x,Q)$ (n = 2,1,0),  so that as $\alpha_s$ increases,
$G(x,Q)\,$will decrease. Second, $\alpha _s$ controls the rate of
evolution of $G(x,Q)$ and hence affects the slope of the gluon
distribution for given measured jet cross-sections. Third, $\alpha
_s(\mu )$ itself depends on $x$ through $\mu =E_t/2=x\sqrt{s}/4$ (at
$\eta =0$), so that the rate of variation of $\alpha _s$ (controlled by
its strength) is coupled to the $x$-dependence of $G(x,Q)$ in the
cross-section formula.

We now apply the results obtained in Sec. \ref{sec:nojet} to these jet data
to see how the latter agree with the predictions of perturbative QCD using
these new parton distributions determined by the other processes.
Fig.~\ref{fig:JetBs}
compares the predictions of the PDF's from the B-series (which
incorporate the most recent DIS data and use the minimal parameters for the
gluon) with the jet data, using the same ``(Data - Theory) / Theory'' format
as Fig.~\ref{fig:JetcM}.
We use the set with $\alpha _s(M_Z)=0.116$ as the
``Theory'' (horizontal solid line) against which the data points as well as
the predictions of the other fits with different $\alpha _s$ values in the
series are displayed in this plot. To make these comparisons, we allow an
overall relative normalization between theory and data.\footnote{%
Such a renormalization, within errors, is usually allowed in global fitting.}
The normalization factor for the CDF/D0 data set ranges from $0.94/0.92$ to $%
1.08/1.06$ for $\alpha _s=0.110$ to $0.122.$ \cite{No105} The normalization
uncertainties quoted by the CDF and D0 experiments are around 5\%. Considering
the 7 orders of magnitude of variation of the cross-section (Fig.~\ref
{fig:JetData}), this is quite remarkable. Within the minimal parametrization
of the gluon used by the B-series, the parton distributions narrowed down
by recent precise DIS data (see previous section) are {\em remarkably
consistent}
with the new high statistics inclusive hadron-hadron jet data.
We also found that the more generally parametrized
C-series PDF's give qualitatively similar predictions for jet cross-sections
compared to the B-series displayed in Fig.~\ref{fig:JetBs}; hence they will
not be separately shown.

The important questions at this point are the following: (i) At a more
quantitative level, how can these parton distributions be improved by
including the jet data in the analysis from the beginning; and (ii)
will the addition of the jet data reduce the variation of $G(x,Q)$
when we use the more general (m+2) parametrization?

\section{New CTEQ parton distribution sets -- CTEQ4}

\label{sec:CTEQ4}

To answer these questions, we have performed an extensive study of the
interplay of the inclusive jet data with the high-precision DIS and other
data within the CTEQ QCD global analysis program. The complete set of
processes and experiments used is given in Table~\ref{tbl:ExptLis}.
To display explicitly the wide coverage of these
experiments over the kinematical variables, we show in Fig.~\ref{fig:KinMap}
a map of the ($x,Q$) plane with the data range of the various experiments.
We see the greatly expanded kinematic coverage compared to a few years ago:
in the direction of small-$x$ due to the HERA experiments, and in the high $%
Q $ direction due to the Tevatron inclusive jet experiments.\footnote{%
Since these experiments are the only ones in the respective kinematic
region, new information on parton distributions extracted from these data
provide challenges to QCD theory for future comparison with independent
measurements based on other processes.} As before, all
processes are treated consistently to NLO accuracy in pQCD. This new round
of global analysis will be referred to as the CTEQ4 analysis.

Building upon studies described in the previous sections, we explored all the
issues described in Sec.~\ref{sec:issues}, now with jet data also playing a
role. Although the quark distributions are coupled to $G(x,Q)$ and $\alpha _s,$
they remain tightly constrained by the DIS experiments, hence stay very close
to those determined before. Thus, our studies concern again mainly the range of
variation of $G(x,Q)$ due to uncertainties in $\alpha _s$ and the
parametrization of the non-perturbative initial distribution. (We have also
looked into the influence due to the choice of ``$Q_{cut}^{}$'', which will be
described in the Appendix).  Since the results from Sec.~\ref{sec:Jets}
indicate that it is possible to obtain good fits to all the data using the
minimal parametrization of the gluon distribution even without taking into
account the experimental systematic errors on the inclusive jet data, we
anticipate the most important role of the latter in the new analysis is to
constrain the possible range of $G(x,Q)$.  Hence, we shall use the more general
(m+2) parametrization which allows a wider range of variation of $G(x,Q)$. We
shall not include the correlated systematic uncertainties on the jet data since
they are not crucial for the present purposes.  This point will come up again
later. More discussions on the experimental systematic uncertainties can be
found in the Appendix.

The new generation of CTEQ4 parton distributions
are summarized in Table~\ref{tbl:CTEQ4PDFs}.
They will be described in turn in the following.

\subsubsection*{Standard CTEQ4M parton distributions}

We first present the standard fit in the $\overline{MS}$ scheme which we
will designate as the CTEQ4M set of parton distributions. The $\alpha
_s(m_Z) $ value for this set is $0.116,$ corresponding to second order $%
\Lambda _5=0.202$ or $\Lambda _4=0.296$ GeV. This set gives excellent fit to
all data sets. The total $\chi ^2$ for $1297$ DIS and DY data points is $1320
$. Detailed information on the $\chi ^2$'s for the various experiments, in
comparison to those obtained using other current and previous generations of
parton distributions are presented in Tables~\ref{tbl:ChiSqA} and \ref
{tbl:ChiSqB} respectively.
The direct photon and jet data sets are not
included in the $\chi ^2$ table since, without including the sizable
theoretical uncertainties for the former\footnote{%
See the Introduction and Refs.~ \cite{CtqDph,jet1} for discussions on these
uncertainties.}~ and experimental systematic errors for the latter, the
significance of such $\chi ^2$ values would be difficult to evaluate. The
comparison of the CDF and D0 jet data to the NLO QCD inclusive jet
cross-section calculated with the CTEQ4M distributions is shown in Fig.~\ref
{fig:JetdM}.
And the comparison of the recent NMC, H1, and ZEUS data sets to
the fit is shown in
Figs.~\ref{fig:DIScdM}.

From Table~\ref{tbl:ChiSqA}, we see that the CTEQ4M PDF set has the best
overall quantitative agreement between NLO QCD theory and global data on
high energy scattering. It also represents a significant improvement over
the previous generation of parton distributions, as a comparison to Table~%
\ref{tbl:ChiSqB} makes clear. Most of the difference is caused by the new
precision data from the HERA experiments. Fig.~\ref{fig:JetdM} shows good
general agreement of CTEQ4M with the jet data, while the much
discussed ``high $E_t^{}$ excess'' is still noticeable.
We will return to this issue in Sec.~\ref{sec:HiJet} where an
alternative ``high $E_t$ jet-fit'' CTEQ4HJ (included in Table~\ref{tbl:ChiSqA})
will be discussed.
Figs.~\ref{fig:DIScdM} explicitly shows the
improvement of CTEQ4M over CTEQ3M in describing the recent high-precision
DIS experiments. In the Appendix, we will give detailed information on the
parameters which characterize the initial parton distributions at $Q_0=1.6$
GeV (which coincides with our choice of the charm threshold). Here, we only
note that the ($A_1,A_2)$ parameters (cf. Eqs.~\ref{iniGlu0} \& \ref{IniGlu1}%
) of the gluon and the sea quarks are $(-1.21,4.67)$ and $(-1.14,8.04)$
respectively.

\subsubsection*{CTEQ4A-series of parton distributions with varying $\alpha
_s $ and $G(x,Q)$}

In exploring the range of variation of allowed $G(x,Q)$ by varying the values
of $\alpha _s,$ changing the number of parameters for the gluon, and altering
the $Q_{cut}$ of data selection, we have found the largest effect is due to the
varying of $\alpha _s.$ Hence, in presenting a series of PDF's which give a
reasonable representation of the range of possibilities, we use those generated
with an $\alpha _s$ range centered around the CTEQ4M value of $0.116,$ which is
close to the current world average \cite{alphasRev}.  This series will be
designated as CTEQ4A-series (shorthand for CTEQ4Alpha)---CTEQ4A1, ...,CTEQ4A5,
with CTEQ4A3 being the same as CTEQ4M.  The $\chi^2$ per point for the 1297
non-jet data points are (1.07,1.02,1.02,1.07,1.19) respectively.  The higher
$\chi ^2$ values at low values of $\alpha _s$ mainly come from the HERA DIS
experiments; the higher $\chi ^2$ values at high values of $\alpha _s$ are
mainly due to the fixed-target DIS experiments \cite{No105}.  The difference in
$\chi^2$ above minimum, especially for the highest value of $\alpha_s$, is
larger than in previous CTEQ analyses (e.g. CTEQ2ML vs. CTEQ2M) due to the
sharply reduced errors on recent DIS data.  However, the difference is
comparable to that between the MRSJ and CTEQ4M $\chi^2s$, cf, Table
\ref{tbl:ChiSqA}.  Because correlations in the experimental errors are not
available for all experiments, hence have not been included in current global
analyses, and since theoretical uncertainties are even harder to quantify,
pragmatically, we take these $\chi^2$ differences as being acceptable for
present purposes.

Fig.~\ref{fig:JetCdA} shows the comparison of the CTEQ4A parton distribution
sets with the two jet data sets, using CTEQ4M as the common calibration.
The overall normalization factor on the jet data sets applied to the various
fits range from $0.96$ (for CTEQ4A1 on D0 points) to $1.02$ (for CTEQ4A5 on
CDF points), well within the experimental uncertainty of $\sim 5\%$.
Comparing to Fig.~\ref{fig:JetBs} and the range of normalization factors
needed there ($0.92-1.08$, which is wider than the experimental error),
we see the expected improvement of the agreement with the jet data.

The gluon distributions associated with the various values of $\alpha _s$ in
this series are shown in Fig.~\ref{fig:GluCdA}. Comparing the CTEQ4A-series
to the C-series (same parametrization form for $G(x,Q_0)$ ), we see that the
constraining influence of the jet data has a rather dramatic effect. The
unstable behavior of the various curves observed in the C-series has been
replaced by an orderly variation as one steps through the values of $\alpha
_s$ within the range explored.
We found, indeed, that for each value of $\alpha _s,$ the solution of $G(x,Q)\,$
is rather unique against perturbations in the fitting procedure.

One concern is that the variation in the CTEQ4A series is too small due to the
lack of treatment of systematic uncertainties in the jet data.  To address this
issue, we compare the change in the calculated jet cross-sections between the
extremes of the CTEQ4A series to the largest $E_t$-dependent uncertainty in the
CDF data. See Fig.~\ref{fig:JetSys}.
It shows that the range of variation in the CTEQ4A series is about 10\% in the
moderate $E_T$ range, while the experimental systematic uncertainty is about
the same.%
\footnote{Of course, given the good agreement between the two Tevatron
experiments \cite{jetnote}, if the CDF jet data requires a significant change
due to a systematic error, the D0 data would require the same change, an
unlikely occurrence since there is almost no correlation in the two
experimental measurements.}  This observation lends some confidence that this
series gives a reasonable estimate of the range of variation of $G(x,Q)$.  To
the extent that there are sources of uncertainty other than $\alpha _s,$ the
variation in $G(x,Q)$ given here may be considered a minimum range. However,
our study does indicate that the variation due to the uncertainty of $\alpha
_s$ may be the dominant one.

Fig.~\ref{fig:xGluSqk} shows a comparison of some of the new gluon and singlet
quark distributions with those of CTEQ3M and MRSJ in the usual form $x\,f(x,Q)$
without the normalization factor as in previous figures. On this conventional
plot, differences in $G(x,Q)$ can be seen only in the small-$x$ region, and the
CTEQ3M and CTEQ4M gluons appear to be indistinguishable.  Differences in the
singlet quark distribution are more evident near $x=0.01$.  The fact that only
small changes in the parton distributions result from adding so much new data
in the global analysis is testament to the impressive progress in pinning down
these parton distributions that has been made in recent years. These changes,
though small, are nonetheless physically significant, as demonstrated by the
substantial differences in $\chi^2$ values between the new and old parton
distribution sets on the precision experiments given in Tables
\ref{tbl:ChiSqA} and~\ref{tbl:ChiSqB}.

\subsubsection*{Other CTEQ4 parton distributions}

Along with the standard CTEQ4M $\overline{MS}$ parton distributions, we have
also obtained corresponding parton distributions in the ``DIS scheme''---
CTEQ4D. CTEQ4D uses the same value of $\alpha _s$ ($=0.116)$ as CTEQ4M; it is
obtained by fitting under identical conditions as CTEQ4M except that the hard
cross-sections are evaluated in the DIS scheme. The $\chi ^2$ values of this
fit are comparable to those of CTEQ4M. In addition to these two standard sets,
for applications requiring leading order (LO) calculations and low values of
the scale $Q$ (LQ), we also provide appropriate parton distribution sets
labelled CTEQ4L and CTEQ4LQ respectively. The CTEQ4LQ set can be used for
$Q_{}^2>Q_i^2=0.5$ GeV$^2.$ 
\footnote{Below this scale ($Q=0.7$ GeV) the QCD coupling approaches unity,
the perturbative formulas certainly cease to be meaningful.}
It was obtained by fitting the same data sets as
the other PDF sets. Since the proper treatment of low $Q$ data must involve
more physics input (such as higher twist effects) than included here, CTEQ4LQ
represents only an extrapolation of twist-two QCD physics into the low $Q$
region---it is not intended to be a best fit.  However, as demonstrated by the
GRV parton distribution sets \cite{GRV}, this kind of extrapolation often turns
out to compare rather well with data in the low $Q$ region.  Comparison of
CTEQ4LQ structure functions to the NMC, E665 and H1 data in the range
$1.0<Q<3.0$ GeV is shown in Fig.~%
\ref{fig:DISLQ}.
The parameters for CTEQ4D, CTEQ4L and CTEQ4LQ are also
given in the Appendix.

The rather remarkably consistent picture resulting from this round of CTEQ4
global analysis incorporating jet data from hadron collisions provides a new
generation of improved parton distributions for making calculations and
predictions on high energy processes both within and beyond the standard
model. The more tightly constrained parton distributions can also lay the
foundation for more stringent tests of the pQCD framework and provide the basis
for discerning signals of new physics.

At present, a remaining area of some uncertainty is the gluon distribution in
the ``large $x$'' region, beyond say $0.25,$ where neither the DIS nor the
direct photon data give tight constraints.  For the DIS process, the
sensitivity to the gluon begins below $x=0.1$.  For the direct photon process
there are a number of theoretical uncertainties which are not yet under
control, as already discussed in the Introduction.  The noticeable rise of the
inclusive jet data points \cite{CdfJets} above all ``theory'' curves shown so
far may be related to the conventional choices of parametrization of the
non-perturbative function $G(x,Q_i)$, which restricts its behavior in the large
$x$ region. This possibility, first raised in Ref.~\cite{jet1}, will be
discussed next in the context of the CTEQ4 analysis presented above.

\section{High $E_t$ Jets and Parton Distributions}

\label{sec:HiJet}

The higher-than-expected inclusive jet cross-sections, first measured by the
 CDF collaboration \cite{CdfJets} for $E_t>200$ GeV, were observed in
 comparison to the existing parton distribution sets, including CTEQ3M as shown
 in Fig.~\ref{fig:JetcM}.  This ``excess'' is reduced slightly when jet data
 are included in the global fit, but is still noticeable in Figs.~\ref
 {fig:JetdM} and \ref{fig:JetCdA} for the CTEQ4A series of distributions.  This
 is understandable since the high $E_t$ data points have large errors, so do
 not carry much statistical weight in the fitting process, and the simple
 (unsigned) $\chi^2$ is not sensitive to the observed pattern that all the
 points are higher than the theoretical prediction in the large $E_t$ region.
 Ref.~\cite{jet1} investigated the feasibility of accommodating these higher
 cross-sections in the conventional QCD framework by exploiting the flexibility
 of $G(x,Q)$ at higher values of $x$ where there are few independent
 constraints, while maintaining the agreement with other data sets in the
 global analysis. To do this, it is necessary to (i) provide enough flexibility
 in the parametrization of $G(x,Q_0)$ to allow for behaviors different from the
 usual (but arbitrary) choice; and (ii) focus on the high $E_t$ data points and
 assign them more statistical weight than their nominal values in order to
 force a better agreement between theory and experiment. Thus, the spirit of
 the investigation is not to obtain a ``best fit'' in the usual sense. Rather,
 it is (i) to find out whether such solutions exist; and (ii) if they do exist,
 to quantify how well these solutions agree with other data sets as compared to
 conventional parton distribution sets. The global analysis work described in
 Sec.~\ref{sec:CTEQ4} without special attention to the high $E_t$ points
 provides the natural setting to put the results of Ref.~\cite{jet1} in
 context.

Ref.~\cite{jet1} was performed using the CDF Run-IA data---the only high
statistics inclusive jet measurement available at the time. Two illustrative
``solutions'' of the type described above were presented---one with the
normalization fixed at 1.0 with respect to the CDF data, the other with a
normalization factor of 0.93. Fig.~\ref{fig:JetHJ} compares predictions of
the normalization=1.0 PDF set, which we shall refer to as the CTEQ4HJ set,
with the more recent Run-IB results of both CDF and D0.
For this comparison,
an overall normalization factor of 1.01(0.98) for the CDF(D0) data set is
found to be optimal in bringing agreement between theory and experiment.%
\footnote{%
The change of CDF normalization factor from $1.0$ to $1.01$ is attributable to
the switch from the Run-IA to the Run-IB data set.} The consistency between the
two data sets, as well as between theory and experiment, displayed by this
comparison appears to be rather remarkable (again, bearing in mind the neglect
of systematic errors other than overall normalization).  Results shown in
Table~\ref{tbl:ChiSqA} quantify the $\chi^2$ values obtained while
accommodating the high $E_t$ jets in the global fit in this particular
case. Compared to the best fit CTEQ4M, the overall $\chi ^2$ for CTEQ4HJ is
indeed slightly higher.  But this difference is much smaller than the
differences discussed earlier in the CTEQ4A series, and much smaller than the
difference between MRSJ and CTEQ4M.  Thus the price for accommodating the high
$E_t$ jets is negligible.  In addition, the difference between CTEQ4HJ and
CTEQ4M is almost entirely due to the BCDMS data, even though the BCDMS $\chi^2$
for CTEQ4HJ by itself is quite good.  This change is due to the fact that, in
the CTEQ4M fit, the BCDMS data set is the dominant one determining the
large-$x$ quark distributions, while, in the CTEQ4HJ fit, the jet data set is
in competition for these quark parameters, and they are changed by minute
amounts.  This is shown in Fig.~\ref{fig:bcdms} where the residuals between
BCDMS data and theory are shown for CTEQ4M and CTEQ4HJ.
The residuals are almost identical, which, together with
Table~\ref{tbl:ChiSqA}, confirms the fact that even though CTEQ4HJ does not
give the absolute overall best fit to all data, it provides an extremely good
description of all data sets. It should be considered as a candidate for the
gluon distribution in nature.\footnote{This is to be contrasted with the
conclusion of {\em incompatibility} between the inclusive jet and DIS data
reached by Ref.~\cite{GMRS}. Their fit to inclusive jet data over the full
$E_t$ range (the MRSJ' set) gives rise to an extremely large $\chi ^2$ for the
BCDMS data set.}  In the future we will need strong, independent measurements
of the large-$x$ gluons in order to clarify the situation with the high-$E_t$
jets.

\section{Summary}

\label{sec:conclude}

In this study of the impact of recent DIS and inclusive
jet data on the global QCD analysis of lepton-hadron and hadron-hadron
processes, we see significant progress in demonstrating the consistency of
the NLO QCD framework, and in narrowing the uncertainties on the elusive but
important gluon distribution. Specifically,

\begin{itemize}
\item  The recent NMC, E665, H1 and ZEUS data considerably narrow down
parton distributions and limit the behavior of the gluon, especially if one
uses the minimal form of the gluon parameterization used by CTEQ3;

\item  The new inclusive jet data agree well with theory predictions based
on PDF's determined by the other processes, with the possible exception of
the high $E_t$ data points.

\item  By adding jet data to the global analysis, it is possible to further
explore the range of variation of the gluon distribution using a more
general parametrization. Although the jet data set covers a limited $x$%
-region, its effect is felt over the entire $x$-range -- because it
complements the other data sets well.

\item  Based on these investigations, a new generation of CTEQ4 parton
distributions for a variety of features are presented: they are tabulated in
Table~\ref{tbl:CTEQ4PDFs}.

\item  Three sources contributing to the uncertainty of the gluon
distribution have been investigated: (i) by letting $\alpha _s$ vary over
its current range of uncertainty; (ii) by increasing the degree of freedom
for parametrizing the non-perturbative initial gluon distribution, and (iii)
by varying the $Q_{cut}$ in selecting data for the global fits. The largest
effect is due to $\alpha_s$.

\item  These studies help to delineate the range of variation of $G(x,Q)$
over the range $10^{-4}<x<0.25$. Further work is needed in exploring
the range of uncertainty of the gluon and other parton distributions by
systematically varying the relevant parameters of the global analysis.

\item  For larger values of $x$, more definitive experimental results on
inclusive jet and direct photon production as well as improved theory are
needed for further progress. The observed high $p_t$ ``excess'' jet
cross-section can be accommodated by a modified gluon distribution,
represented by the CTEQ4HJ set, since no other independent measurement
constrains it in this range.

\end{itemize}

In view of the strong correlation between the gluon distribution and $\alpha
_s,$ narrowing the uncertainty in the latter will significantly improve the
determination of $G(x,Q).$ What can a global analysis of experimental data
described in this paper contribute to the measurement of $\alpha _s?$ To
explore this question, one needs to study in some detail the sensitivity of
each process which contributes to the global analysis to the variation of $%
\alpha _s$. This problem will be pursued in a separate analysis.

\section*{Appendix}

\label{sec:App}

\subsubsection*{CTEQ4 Parton Distribution Parameters}

The initial parton distributions at $Q=Q_0,\,f^i(x,Q_0),$ are parametrized
in general as in Eq.~\ref{IniGlu2} for the gluon $G$ and the quark flavors $%
d_v,\,u_v,\,\bar{u}+\bar{d},\,s\,(\bar{s});$ except for the combination $%
\bar{d}-\bar{u}$ (which does not have to be positive definite) which is
parametrized as:
\[
\bar{d}-\bar{u}=A_0\,x^{A_1}\,(1-x)^{A_2}\,(1+A_3\sqrt{x}+A_4\,x)
\]
For all parton distribution sets, $Q_0=1.6$ GeV, except for CTEQ4LQ which has
$Q_0=0.7$ GeV. Tables of the coefficients \{$A_n^i;\ n=1,..,4;\ i=flavors\}$ for
the three standard parton distribution sets CTEQ4M, CTEQ4D, CTEQ4L and the
low-$Q_0$ set CTEQ4LQ are given
below, in Tables~\ref{tbl:Mparam},\ref{tbl:Dparam}, \ref{tbl:Lparam} and 
\ref{tbl:LQparam}.
All parton
distribution sets listed in Table~\ref{tbl:CTEQ4PDFs} are available in
fortran program form by request \footnote{%
Requests can be sent to Lai\_H@Pa.Msu.Edu or Tung@Pa.Msu.Edu.} or via WWW at
http://www.phys.psu.edu/\~{}cteq/.

\subsubsection*{Experimental Normalization Factors}

The $\chi ^2$ tables \ref{tbl:ChiSqA},\ref{tbl:ChiSqB} are obtained by allowing
the experimental data sets to ``float'' with respect to the theory
cross-sections. For CTEQ distributions, a $\chi ^2$ penalty is included in
the fitting process for deviations of the normalization factors with respect
to the respective overall experimental normalization errors. For non-CTEQ
distributions, we simply obtained the minimum $\chi ^2$ by varying the
normalization factors without such penalty. The resulting normalization
factors which go with Tables \ref{tbl:ChiSqA},\ref{tbl:ChiSqB} are given in 
Tables~\ref{tbl:ExpNorA} and \ref{tbl:ExpNorB}.

\subsubsection*{Experimental Systematic Uncertainties}

For DIS, DY and direct photon data, we follow the usual procedure of combining
in quadrature point-to-point systematic errors given by the experiments
with the statistical errors. Correlated systematic errors other than
overall normalization are not generally available from most experiments.
For a few where they are, we have done separate studies of the
consequences of incorporating them in the global analysis and found they
do not affect the best fit parameters by any significant amount.
See Ref.~\cite{ctqlng}.

For the preliminary inclusive jet data, only the normalization
uncertainty is taken into account in the global fit. The rationale
has been explained in Sec.~\ref{sec:CTEQ4}.
The fully correlated systematic errors from CDF,
although available, are not easily implemented in a way
which is consistent with all the other data sets.
(A separate study on the effects of these uncertainties employing the
full correlation matrix is underway,
and it will be reported in the future.)
These errors are not yet available for the D0 data set.
The main effect of omitting the systematic errors on jets is to increase
somewhat the relative weight of this data set in the global analysis. This
will not affect the fits substantially because the jet data agree well with
parton distributions determined from other processes, as discussed in
Sec.~\ref{sec:Jets}.

In general, the
question of assigning appropriate relative weights to different experimental
data sets in a global analysis is a difficult one. An experiment with few
data points which is however particularly sensitive to some physical
parameters than all the others can sometimes be emphasized justifiably in a
global $\chi ^2$ minimization process, otherwise it will be overwhelmed by
the far more numerous data sets and the sensitivity will be lost. As an
extreme example, the NA51 experiment \cite{NA51}, which has an important
impact on the determination of the flavor SU(2) assymmetry of the sea quarks
($\bar{u}-\bar{d}$), consists of only one data point. It has to be
appropriately emphasized in a global analysis to have an effect in
differentiating the sea quarks.

\subsubsection*{Dependence on the Choice of $Q_{cut}$}

\label{sec:qcut}

In all global QCD studies, a set of cut-offs on the hard scale ``Q'' for
various processes is used in data selection. In recent CTEQ analyses, this $%
Q_{cut}$ has been 2 GeV on $Q$ and 3.5 GeV on $W$ in DIS, 2 GeV on $Q$ (the
invariant lepton pair mass) in Drell-Yan process, and 4 GeV on $p_t$ in
direct photon production. As a final check on the reliability of the results
described in the previous section, we test the sensitivity of the fits to
the value of these cut-offs in order to gauge possible influence due to
non-perturbative or higher-twist effects.\footnote{%
This issue has previously been investigated in Ref.~\cite{MorTun}. The
accuracy of both experiments and theory have improved dramatically since
then.} For this purpose, we carried out several series of analyses similar
to the CTEQ4A-series above, but with the minimum $Q_{cut}$ raised progressively
from 2 GeV to 3, 4, and 5 GeV. Data points excluded by these higher $Q_{cut}$%
's are mainly those of fixed-target DIS experiments. We found our results to
be rather stable under these changes. Fig.~\ref{fig:QcutGlu} compares the
gluon distributions from three PDF sets obtained with three $Q_{cut}$ values
mentioned above, all for a given $\alpha _s$ value of 0.113.
We see that the
differences are quite small -- smaller than those due to the variation of $%
\alpha _s$ (with the same $Q_{cut}$) shown in Fig.~\ref{fig:JetCdA}
and described in Sec.~\ref{sec:CTEQ4}.
The subtle differences, especially in relation to sensitivity on $\alpha _s$
values, will be discussed elsewhere \cite{CtqAls}.

\subsubsection*{Acknowledgement}
We would like to thank our CTEQ colleagues, R. Brock, J. Collins, J. Morfin,
J. Pumplin, J.W. Qiu, J. Smith, G. Sterman, J. Whitmore, and C.-P. Yuan,
for very useful discussions and encouragement.


\newtheorem{Figure}{Figure}

\newcommand{\capt}[1]{\rm #1}

\input epsf

\def\figNmcHacM
{
\begin{Figure}
\epsfysize=3.2in
\centerline{\epsfbox{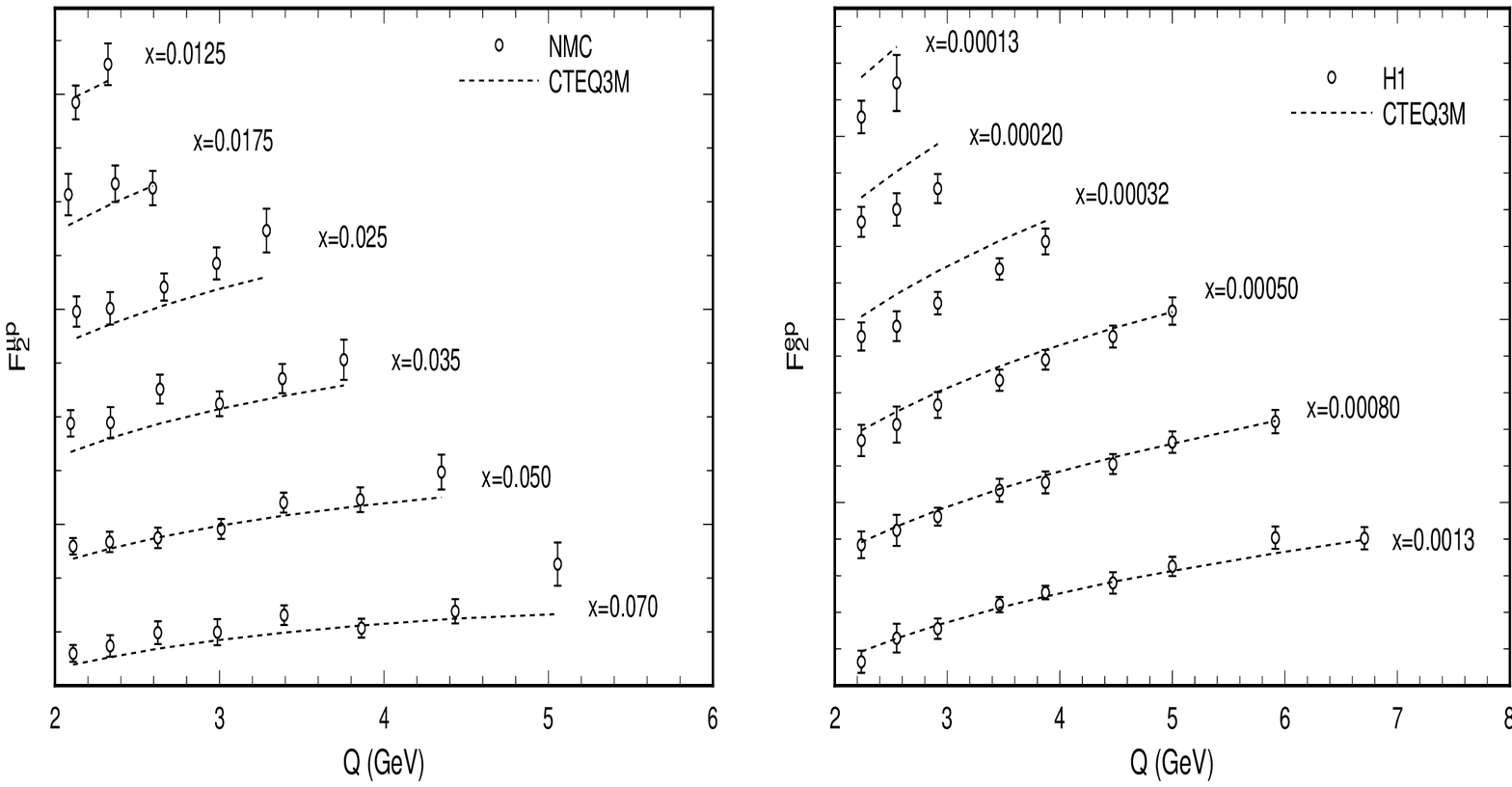}}

  \capt{: Comparison of NLO calculations based on the previous
generation CTEQ3M parton distributions with the latest NMC (a) and H1 (b)
data in the small-$x$ region where discrepencies appear.}
  \label{fig:NmcHacM}
\end{Figure}
}
\def\figGluAa
{
\begin{Figure}
\epsfysize=8in
\centerline{\epsfbox{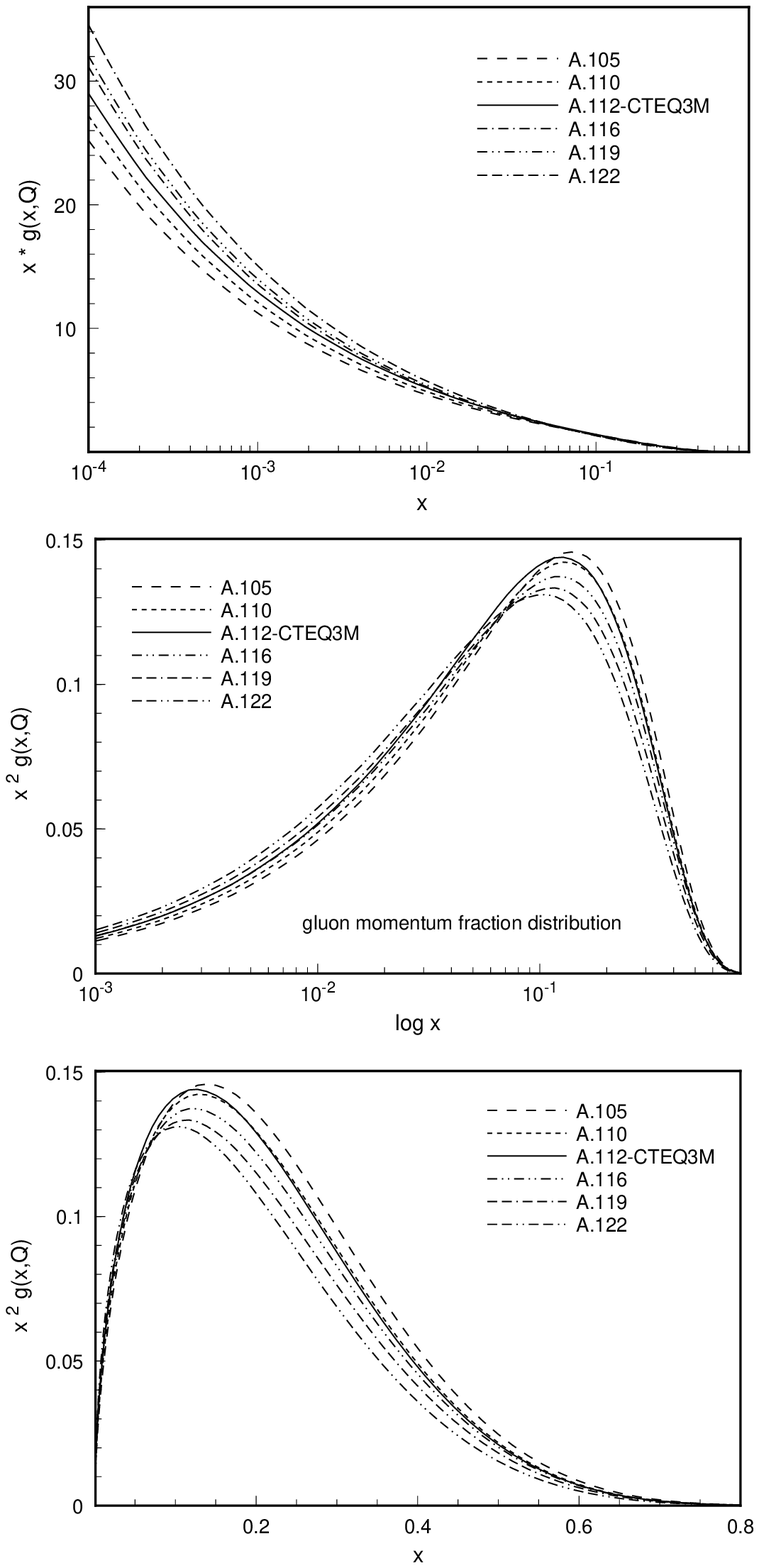}}

  \capt{: Series-A gluon distributions in the small-, medium-, and
large-$x$ regions. A.105 refers to the gluon associated with
$\alpha_s(M_Z)=0.105$, and likewise for the other ones.}
  \label{fig:GluAa}
\end{Figure}
}
\def\figGluA
{
\begin{Figure}
\epsfysize=3.2in
\centerline{\epsfbox{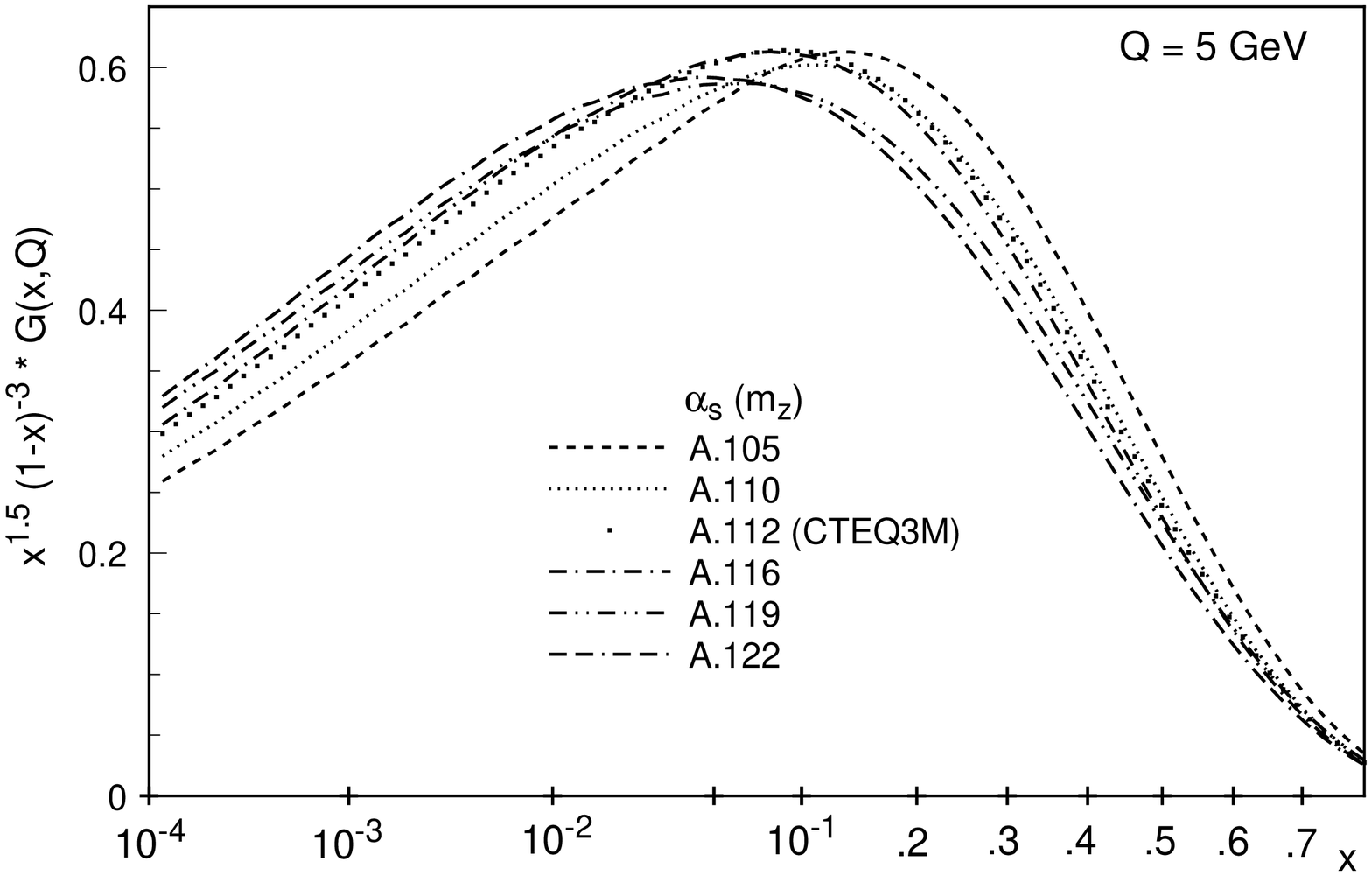}}

  \capt{: Series-A gluon distributions normalized by the function
$x^{-1.5}(1-x)^3$ in order to display clearly the behavior
of $G(x,Q)$ over the entire $x$-range. For the same purpose, the horizontal
x-axis is drawn with a scale which smoothly
changes from log- to linear behavior. }
  \label{fig:GluA}
\end{Figure}
}

\def\figGluB
{
\begin{Figure}
\epsfysize=3.2in
\centerline{\epsfbox{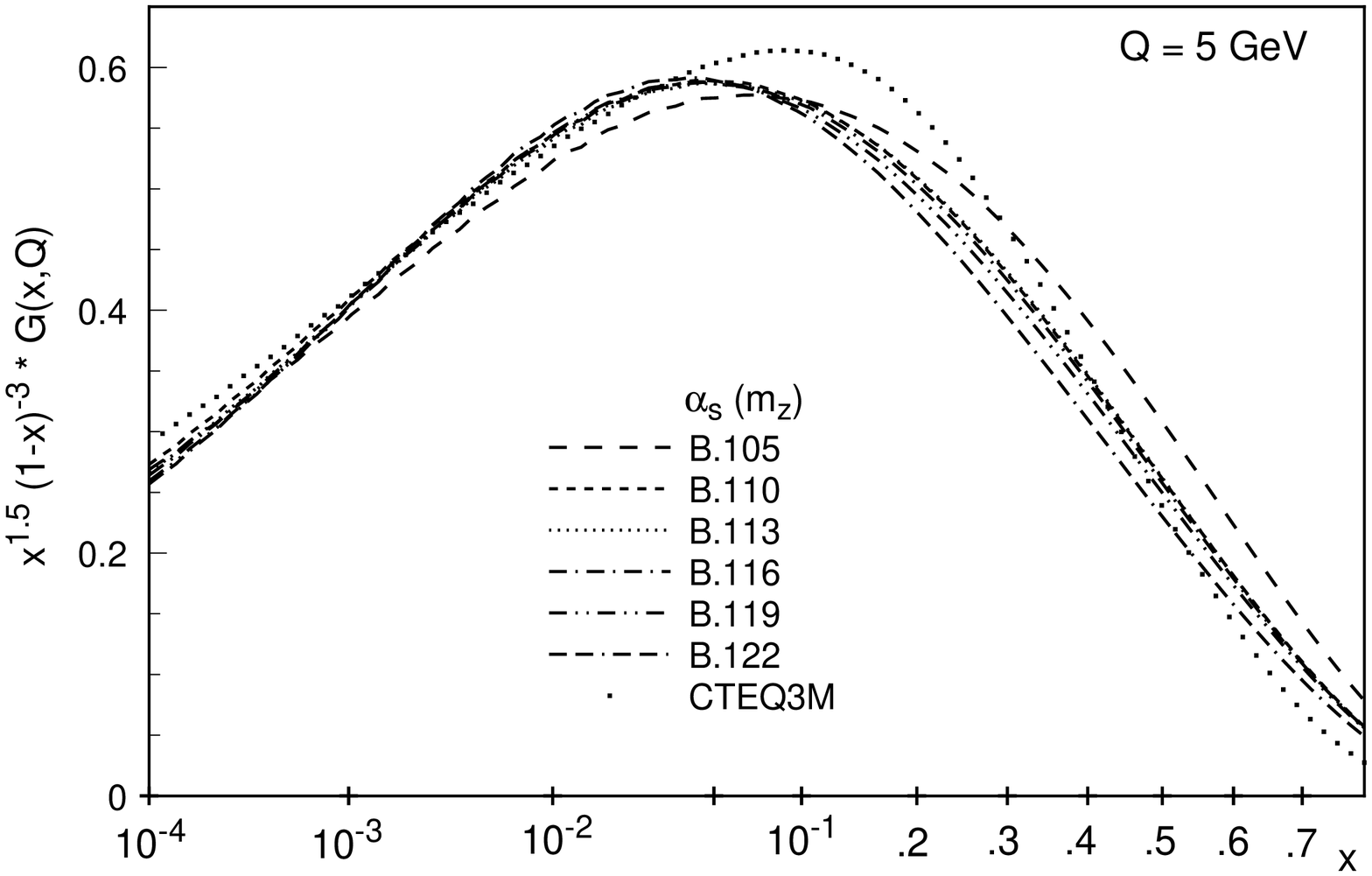}}

  \capt{: Series-B gluon distributions normalized by the function
$x^{-1.5}(1-x)^3$ (cf. caption of previous figure.)}
  \label{fig:GluB}
\end{Figure}
}

\def\figGluC
{
\begin{Figure}
\epsfysize=3.2in
\centerline{\epsfbox{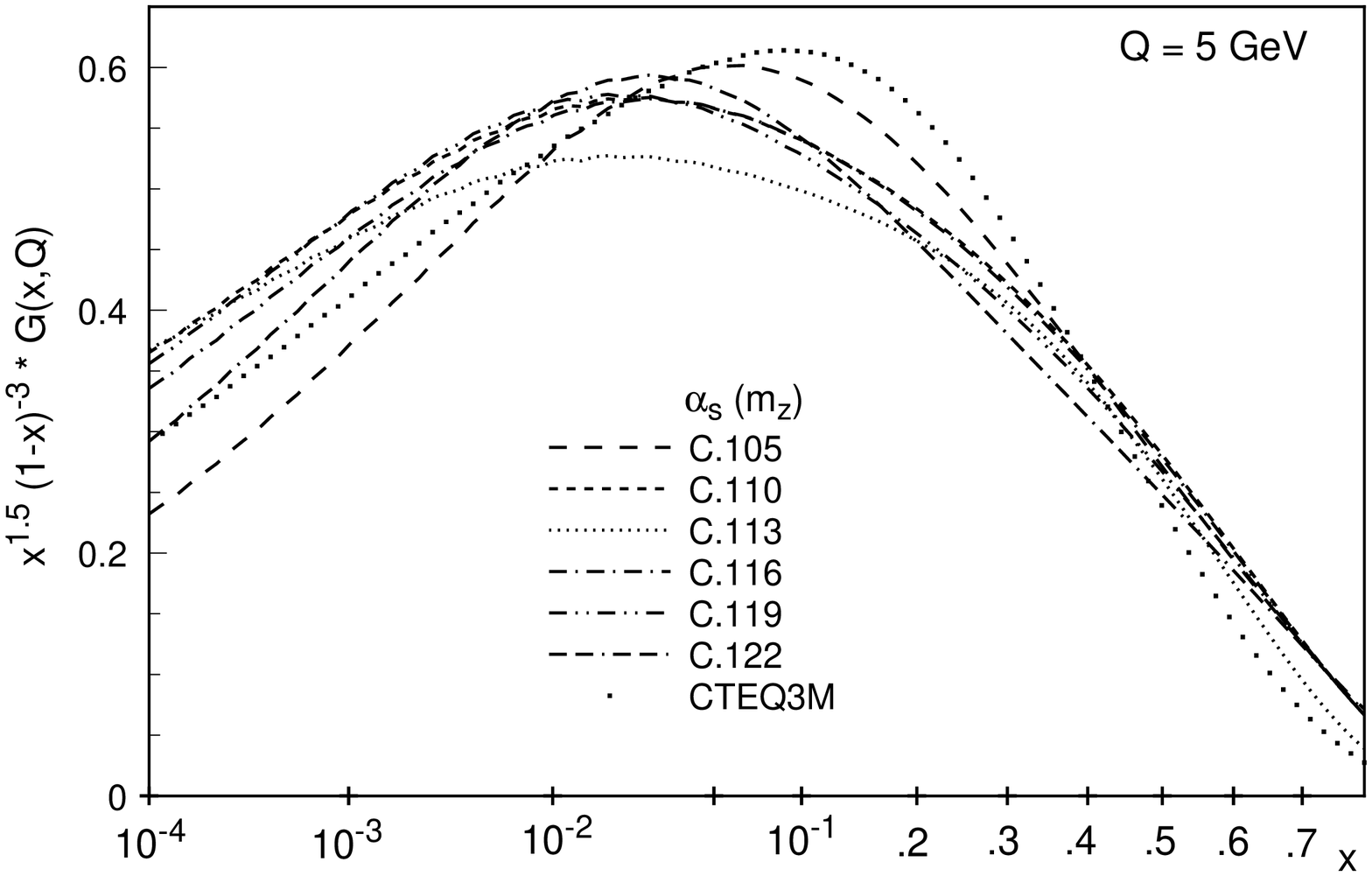}}

  \capt{: Series-C gluon distributions normalized by the function
$x^{-1.5}(1-x)^3$ }
  \label{fig:GluC}
\end{Figure}
}

\def\figJetData
{
\begin{Figure}
\epsfysize=3.2in
\centerline{\epsfbox{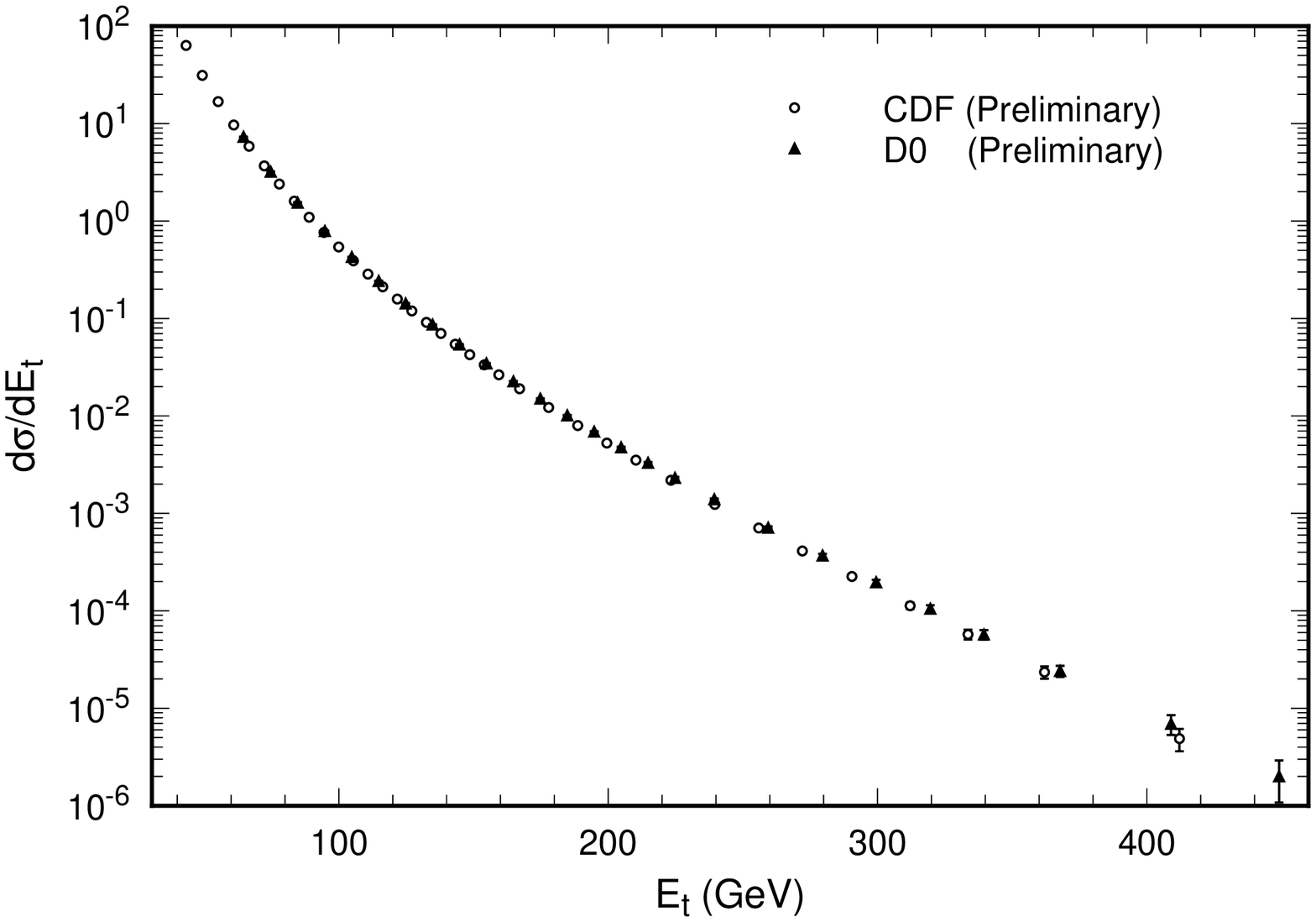}}

  \capt{: Inclusive jet cross-section measured by the CDF and D0
collaborations in Run-IB at the Tevatron.
(Averaged over $0.1 < |\eta| < 0.7$ in the case of CDF and
$|\eta|<0.5$ in the case of D0.)
}
  \label{fig:JetData}
\end{Figure}
}

\def\figJetUncer
{
\begin{Figure}
\epsfysize=3.2in
\centerline{\epsfbox{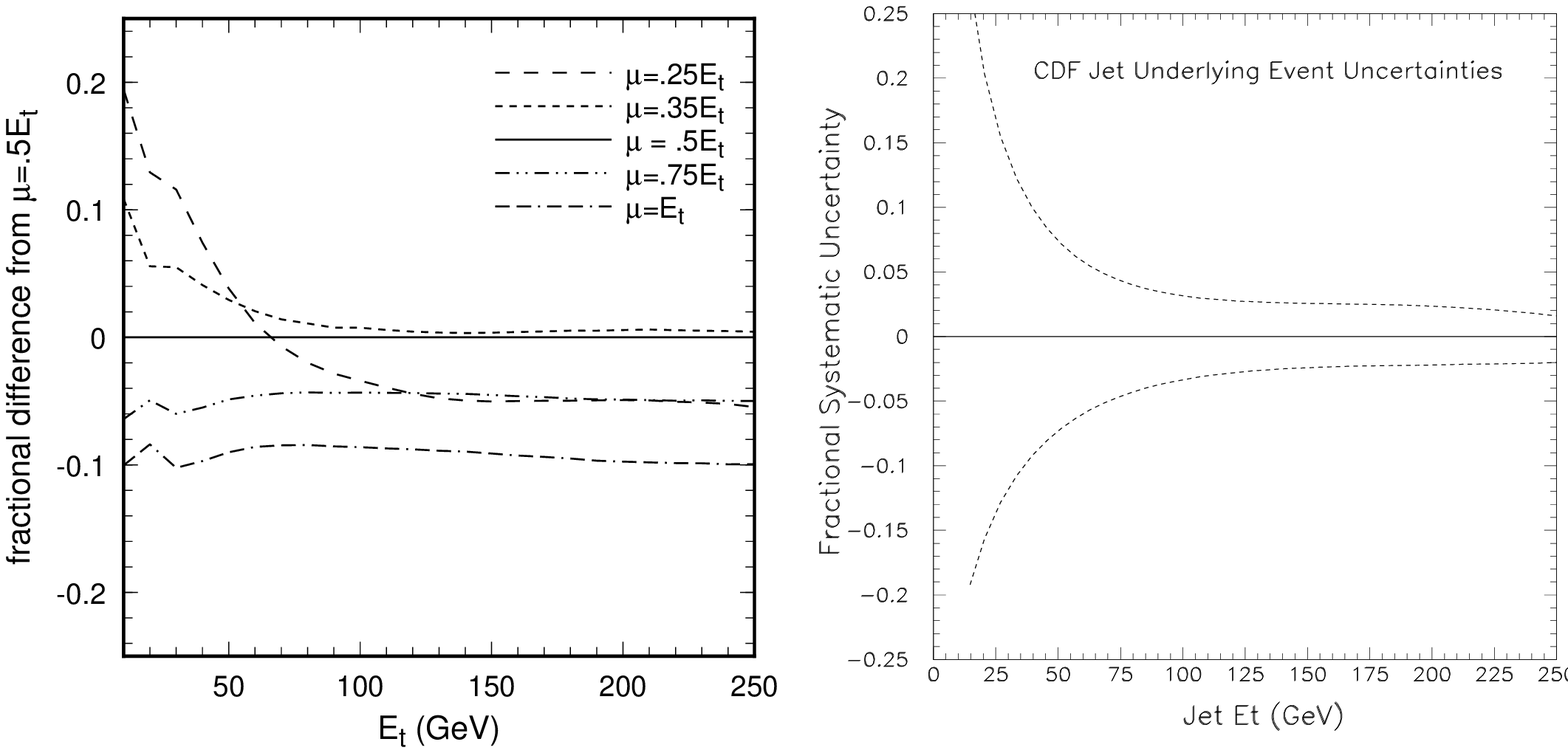}}

  \capt{: Two examples of sources of uncertainties in comparing inclusive
jet data with NLO QCD theory: (a) Fractional difference between
${d\sigma }(E_t,\mu)/{dE_t}$ and  ${d\sigma }(E_t,\mu=E_t/2)/{dE_t}$
(for the CDF rapidity coverage $0.1<|\eta|<0.7$)
as a function of $E_t$ for a variety values of $\mu$;
(b) Fractional change in the cross-section due to $\pm 30$ \% change
in underlying event correction in the CDF experiment.}
  \label{fig:JetUncer}\end{Figure}
}

\def\figJetcM
{
\begin{Figure}
\epsfysize=3.2in
\centerline{\epsfbox{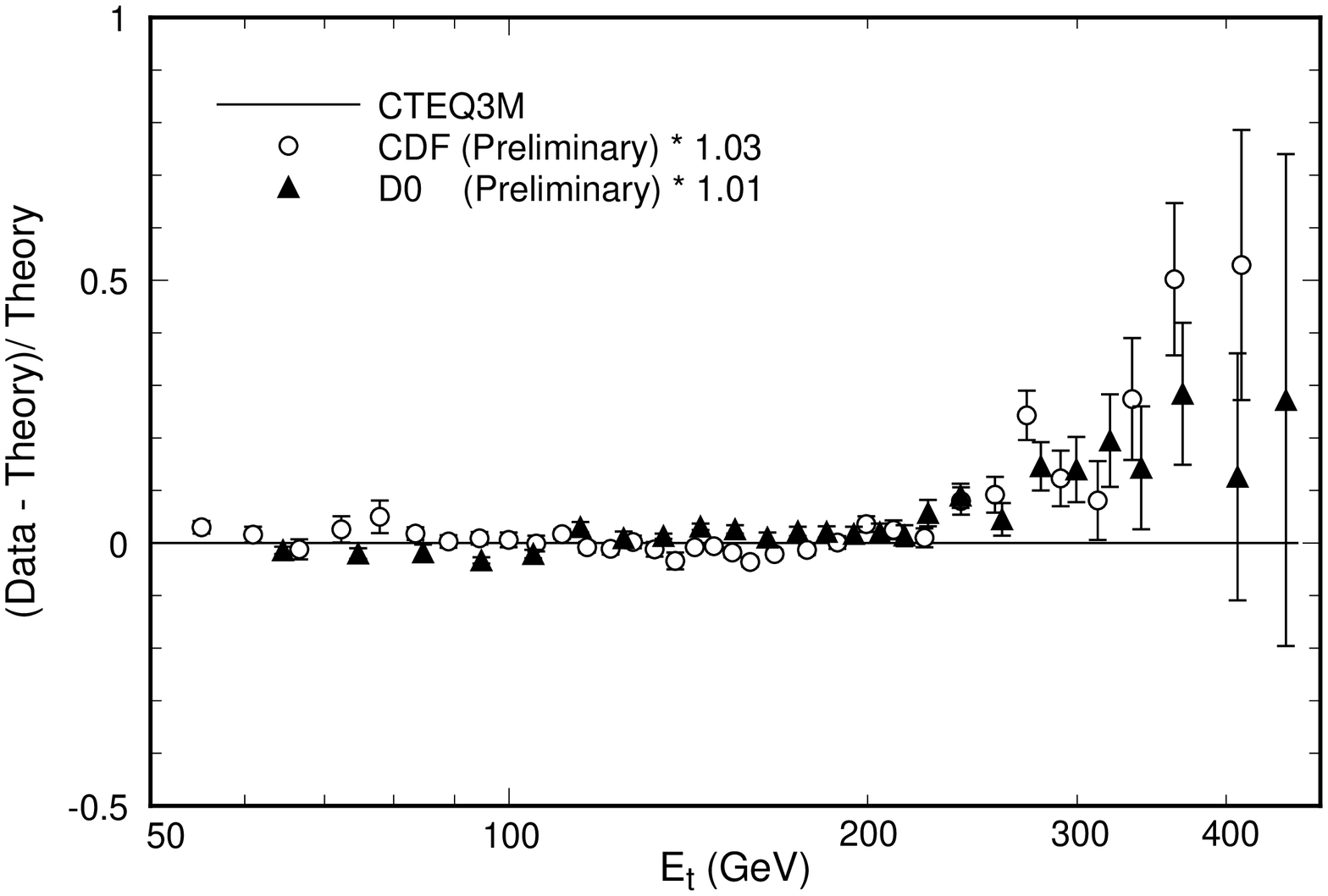}}

  \capt{: Inclusive jet cross-section measured by the CDF and D0
collaborations in Run-IB at the Tevatron normalized to NLO QCD calculations
based on CTEQ3M PDF's. The difference in rapidity coverage of the two
experiments is taken into account. }
  \label{fig:JetcM}
\end{Figure}
}

\def\figSubProc
{
\begin{Figure}
\epsfysize=4.3in
\centerline{\epsfbox{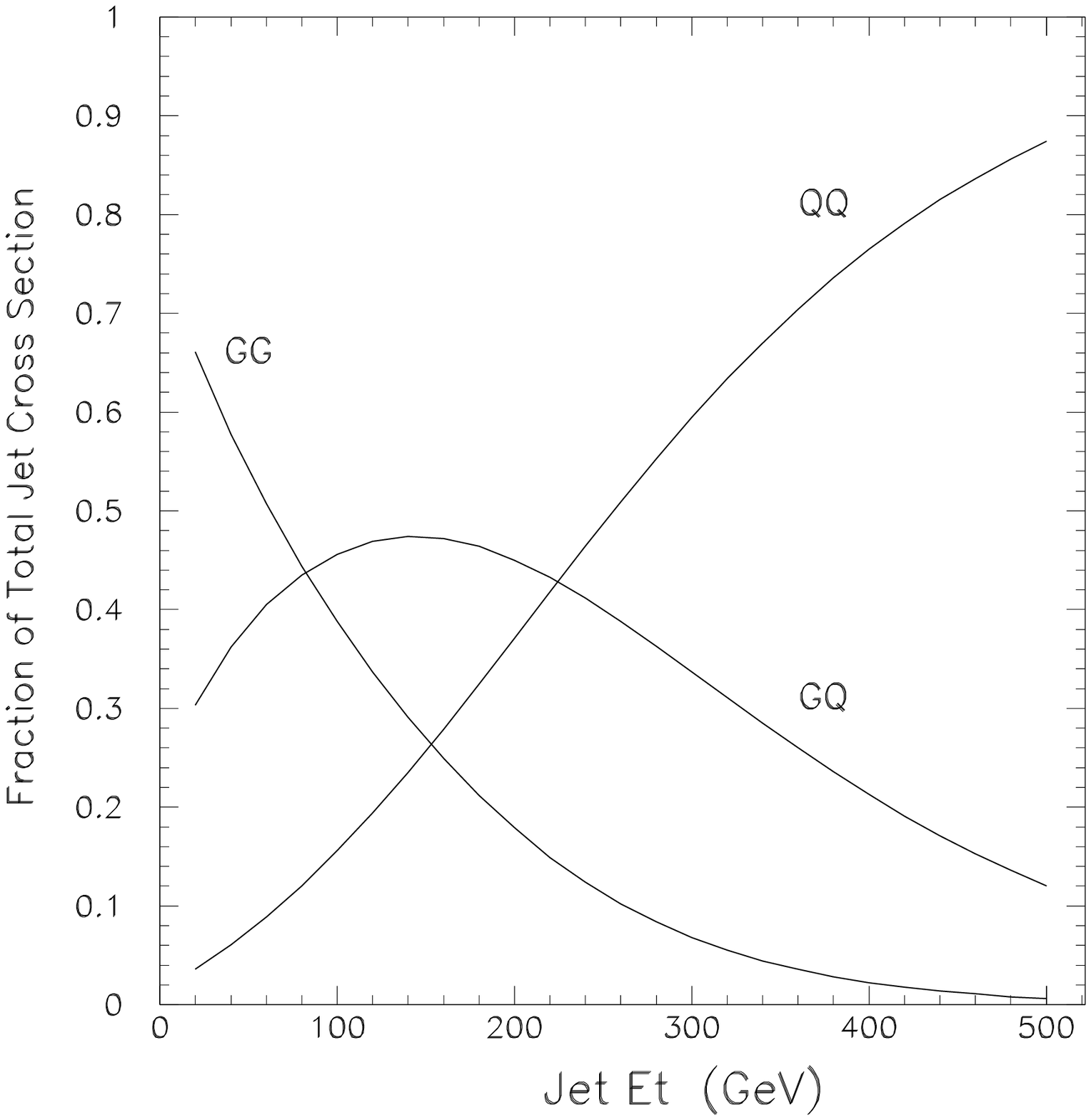}}
 \capt{: Relative contribution to the inclusive jet cross-section due to the
various partonic subprocesses.}
  \label{fig:SubProc}
\end{Figure}
}

\def\figJetBs
{
\begin{Figure}
\epsfysize=3.2in
\centerline{\epsfbox{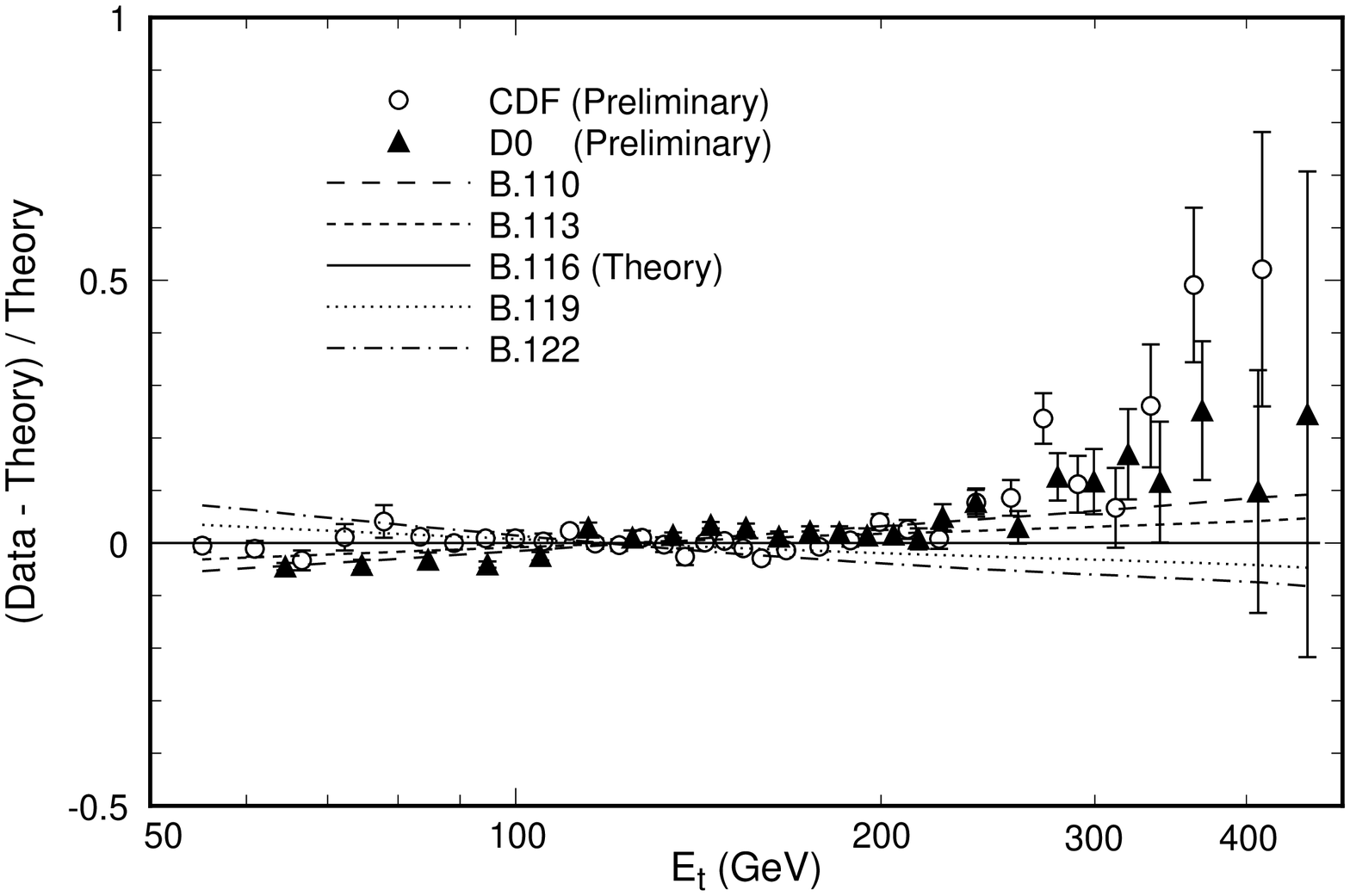}}

  \capt{: Inclusive jet cross-section of CDF and D0
compared to NLO QCD calculations based on the new B-series parton
distributions.}
  \label{fig:JetBs}
\end{Figure}
}

\def\figKinMap
{
\begin{Figure}
\epsfysize=4in
\centerline{\epsfbox{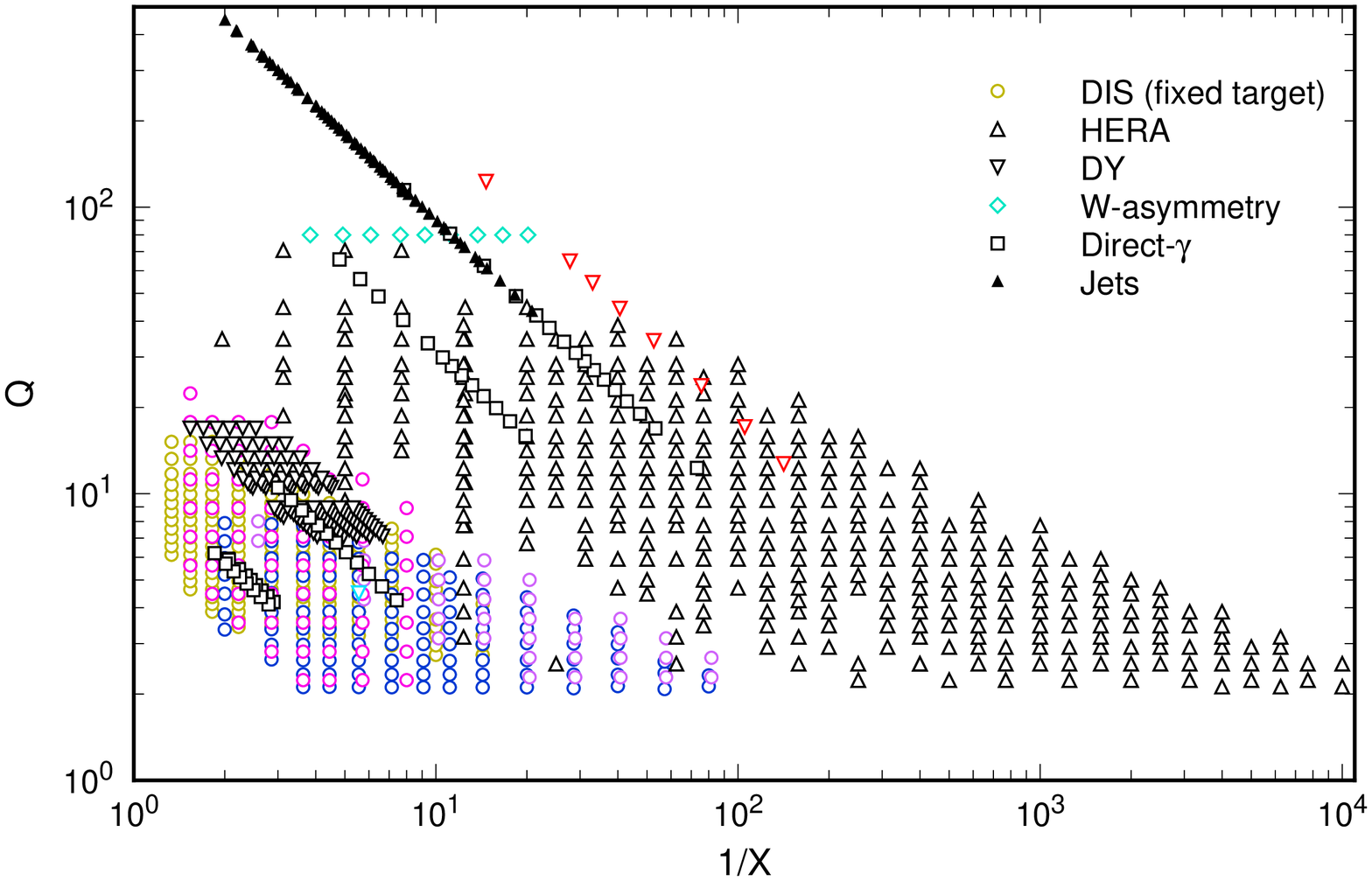}}

  \capt{: Kinematic map in the $(x,Q)$ plane of data points used in the
current global analysis.}
  \label{fig:KinMap}
\end{Figure}
}

\def\figJetdM
{
\begin{Figure}
\epsfysize=3.2in
\centerline{\epsfbox{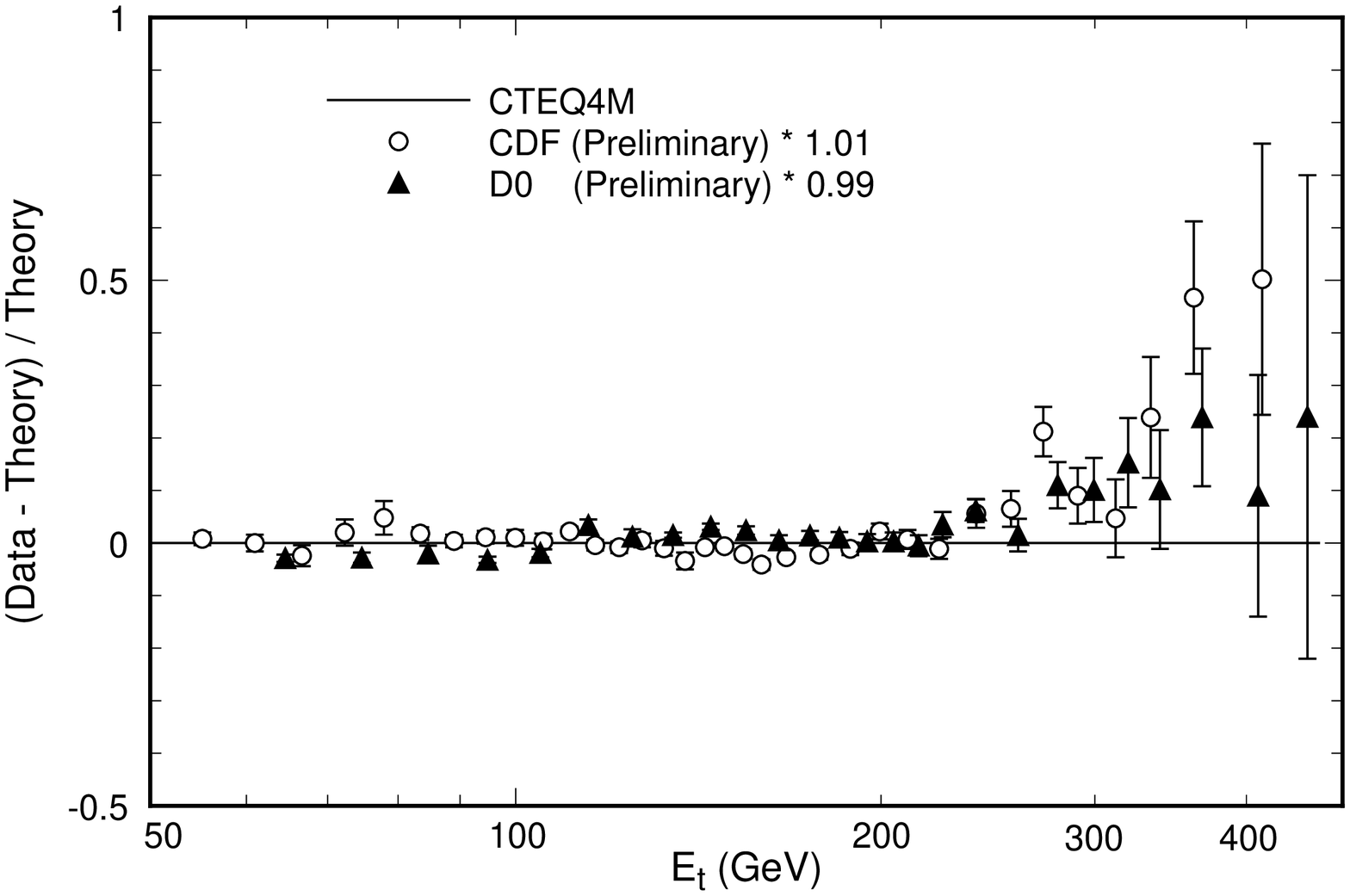}}

  \capt{: Inclusive jet cross-section of CDF and D0
compared to NLO QCD calculations based on the new CTEQ4M parton
distributions.}
  \label{fig:JetdM}
\end{Figure}
}

\def\figDIScdM
{
\begin{Figure}
\epsfysize=8in
\centerline{\epsfbox{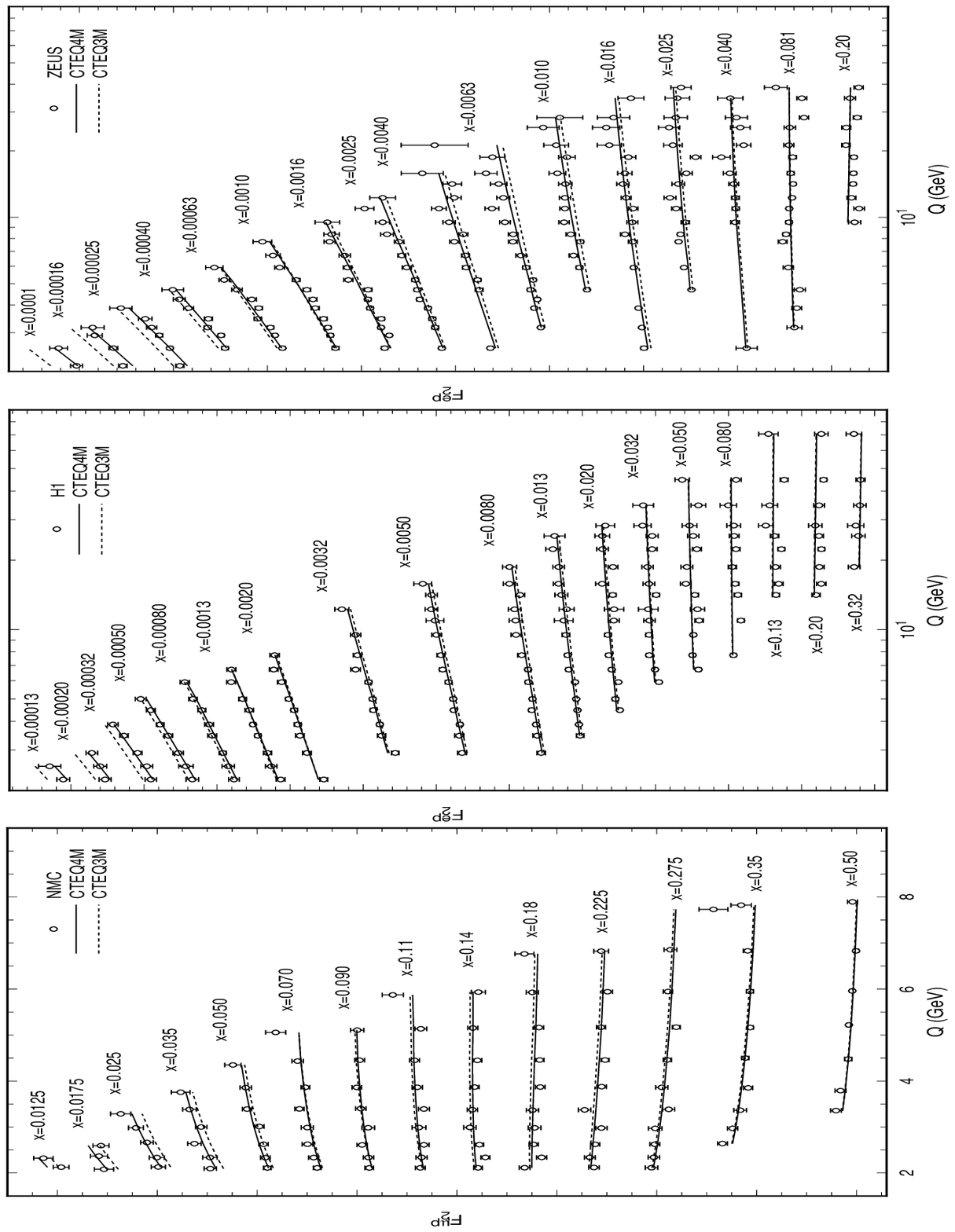}}

  \capt{: Comparison of $F_2^{p}$ data from NMC, H1 and ZEUS to
NLO QCD calculations based on CTEQ3M and CTEQ4M.
The improvement in the small-$x$ region is evident.}
  \label{fig:DIScdM}\end{Figure}
}

\def\figJetCdA
{
\begin{Figure}
\epsfysize=3.2in
\centerline{\epsfbox{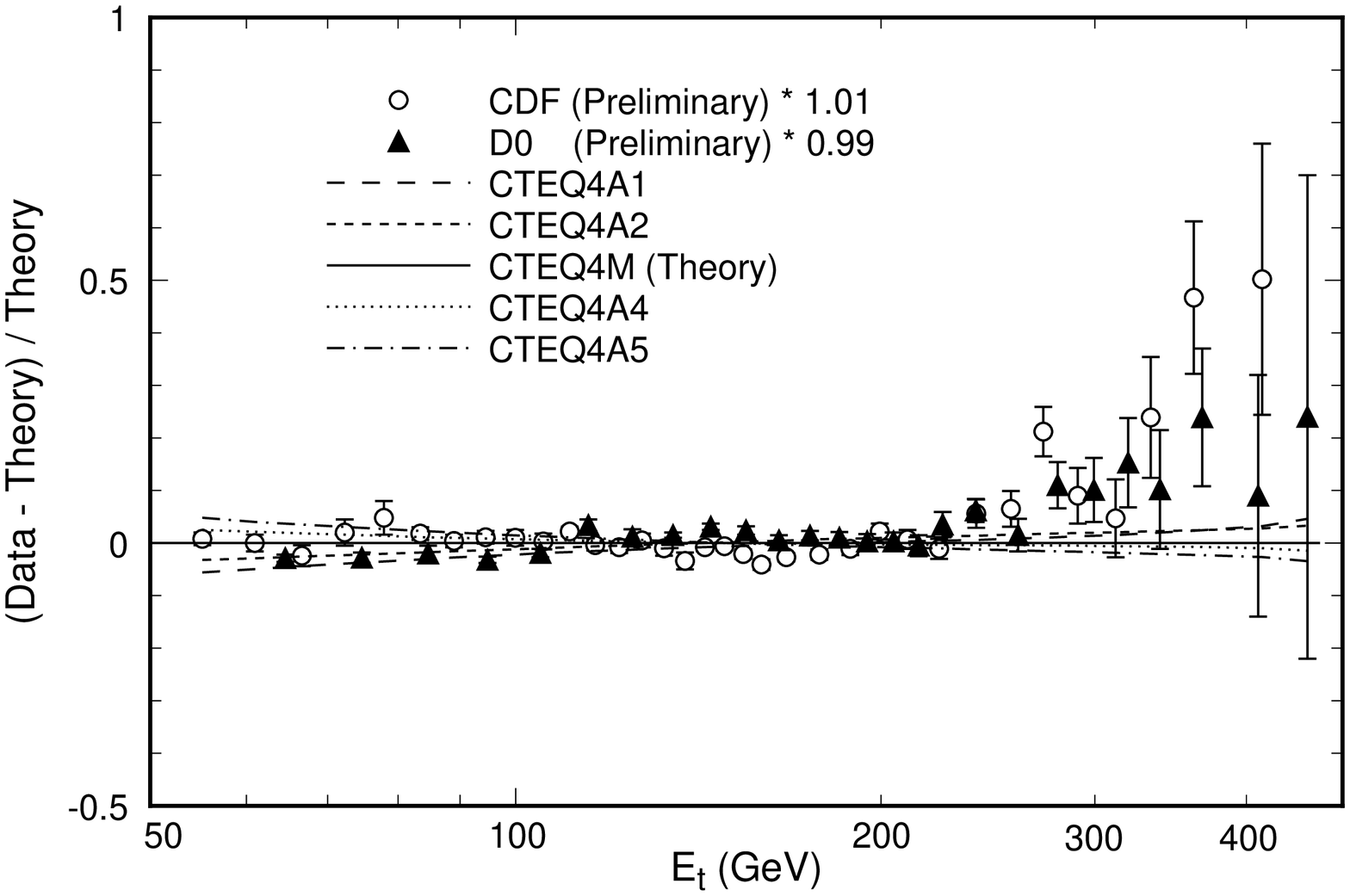}}

  \capt{: Inclusive jet cross-section of CDF and D0
compared to NLO QCD calculations based on the new CTEQ4A series of parton
distributions.}
  \label{fig:JetCdA}
\end{Figure}
}

\def\figGluCdA
{
\begin{Figure}
\epsfysize=3.2in
\centerline{\epsfbox{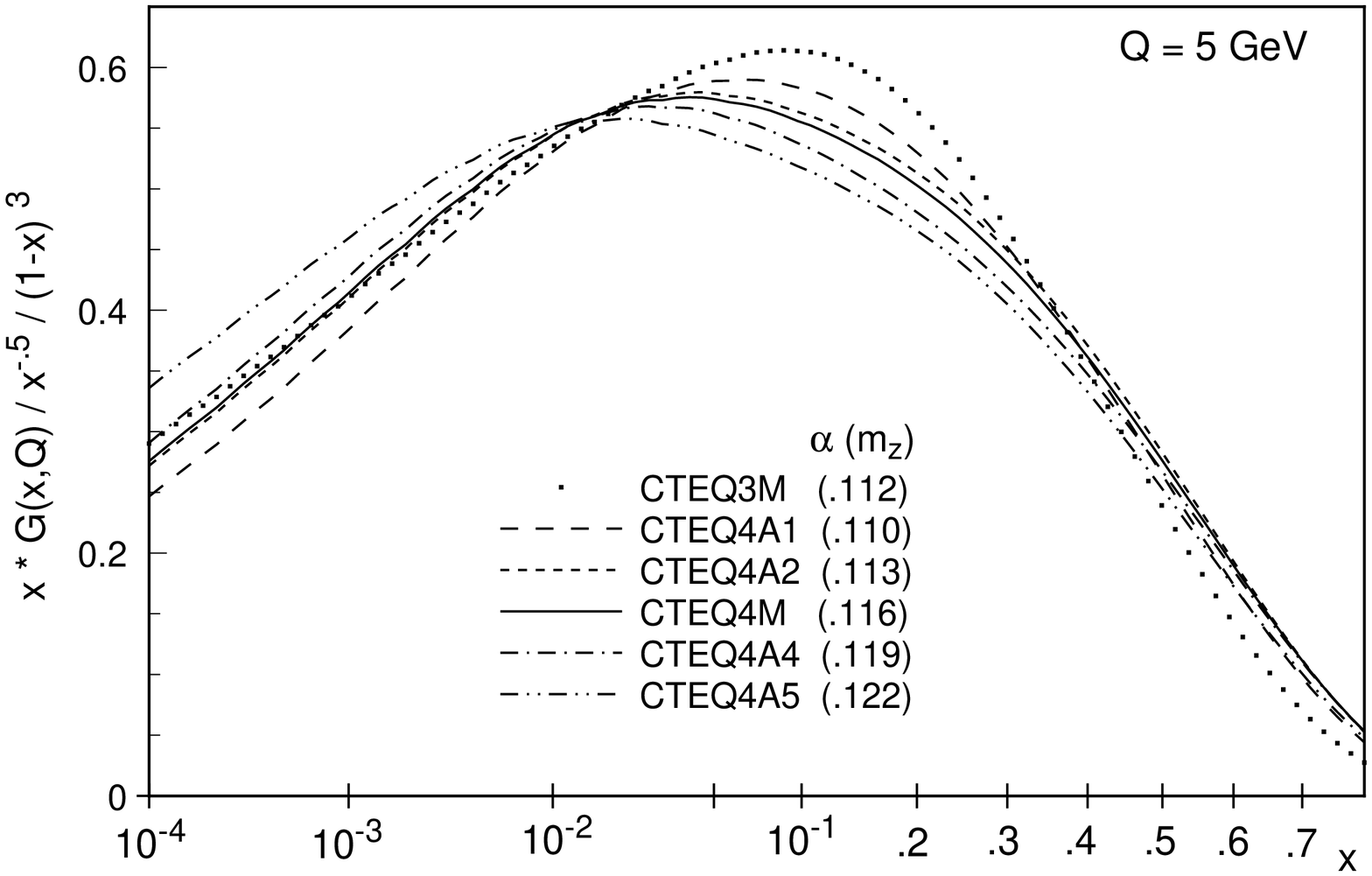}}

  \capt{: Series-CTEQ4A gluon distributions normalized by the function
$x^{-1.5}(1-x)^3$.}
  \label{fig:GluCdA}
\end{Figure}
}

\def\figJetSys
{
\begin{Figure}
\epsfysize=3.2in
\centerline{\epsfbox{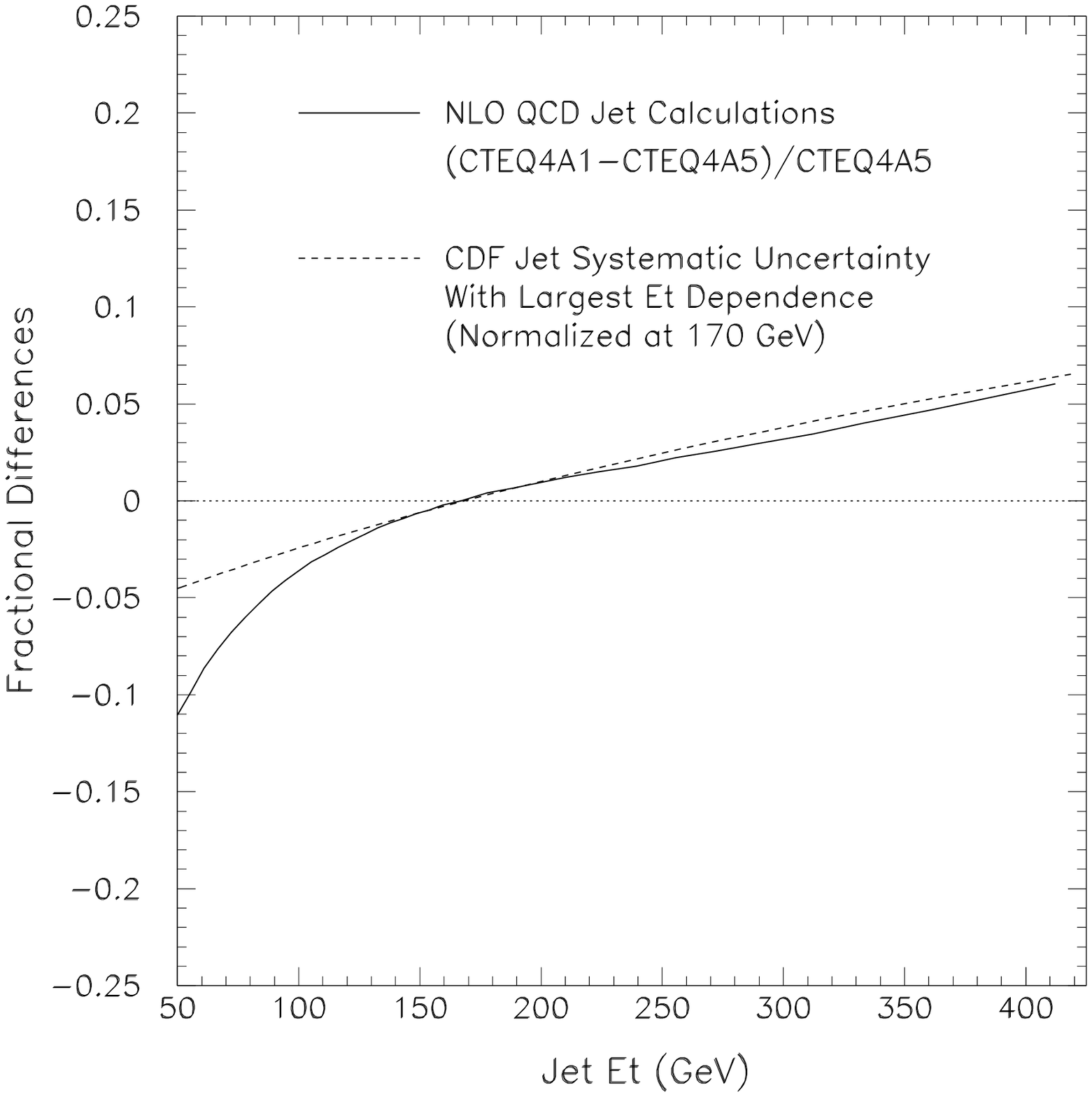}}

  \capt{: Percentage range of variation of the inclusive jet cross-section
from the two extreme CTEQ4A PDF sets (CTEQ4A1 and CTEQ4A5) compared to the
largest of the $E_t$ dependent systematic uncertainties.}
  \label{fig:JetSys}
\end{Figure}
}

\def\figDISLQ
{
\begin{Figure}
\epsfysize=7in
\centerline{\epsfbox{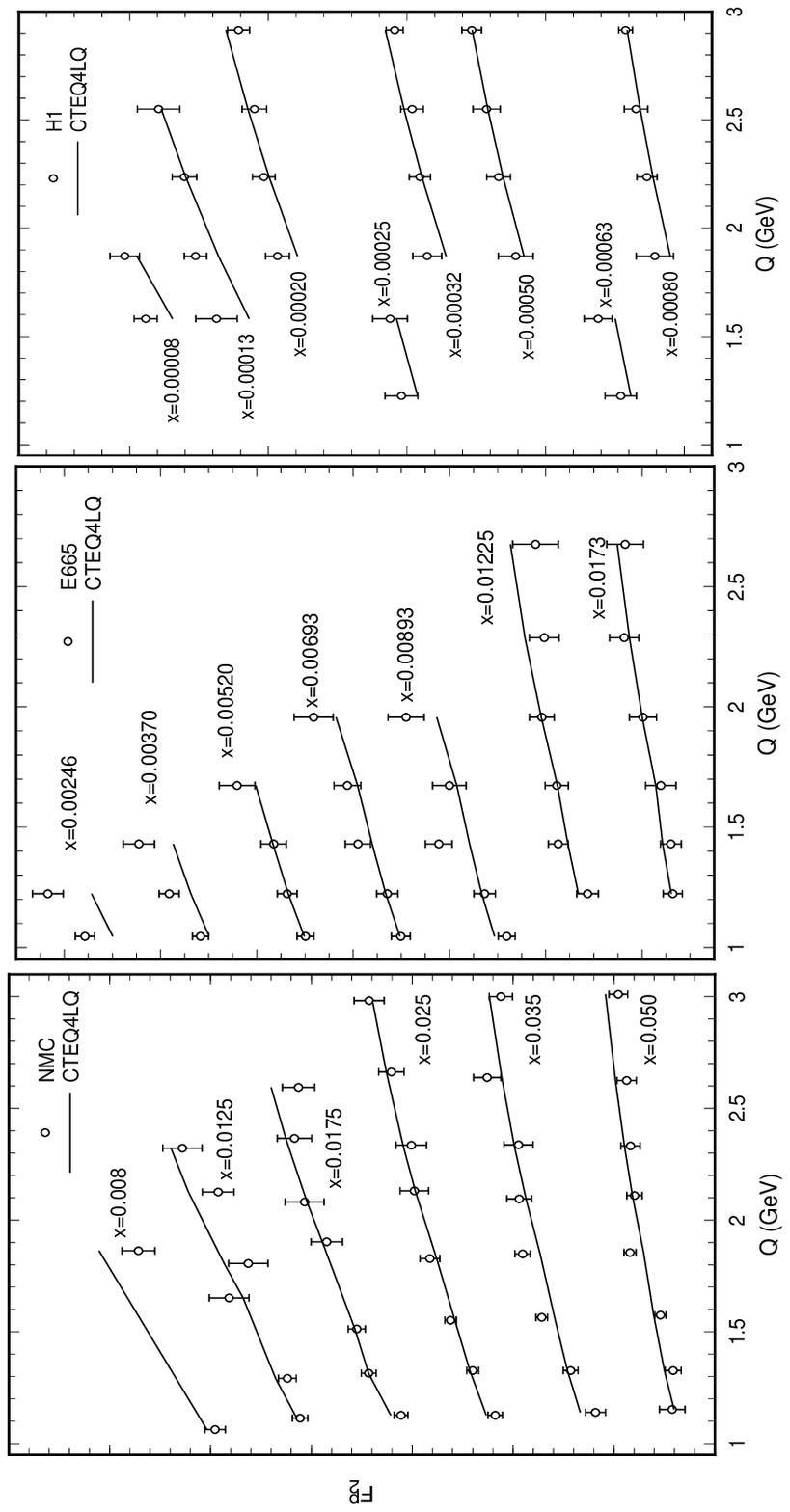}}

  \capt{: Comparison of $F_2^{p}$ data in the low-$Q$ region from H1, E665 and
NMC to NLO QCD calculations based on CTEQ4LQ PDF's. CTEQ4LQ is obtained by
fitting to data with $Q>2$ GeV only. The extrapolation to below $Q=2$ GeV
appears to work remarkably well except for the two lowest $x$ bins of the E665
data shown.}  \label{fig:DISLQ}
\end{Figure}
}

\def\figQcutGlu
{
\begin{Figure}
\epsfysize=3.2in
\centerline{\epsfbox{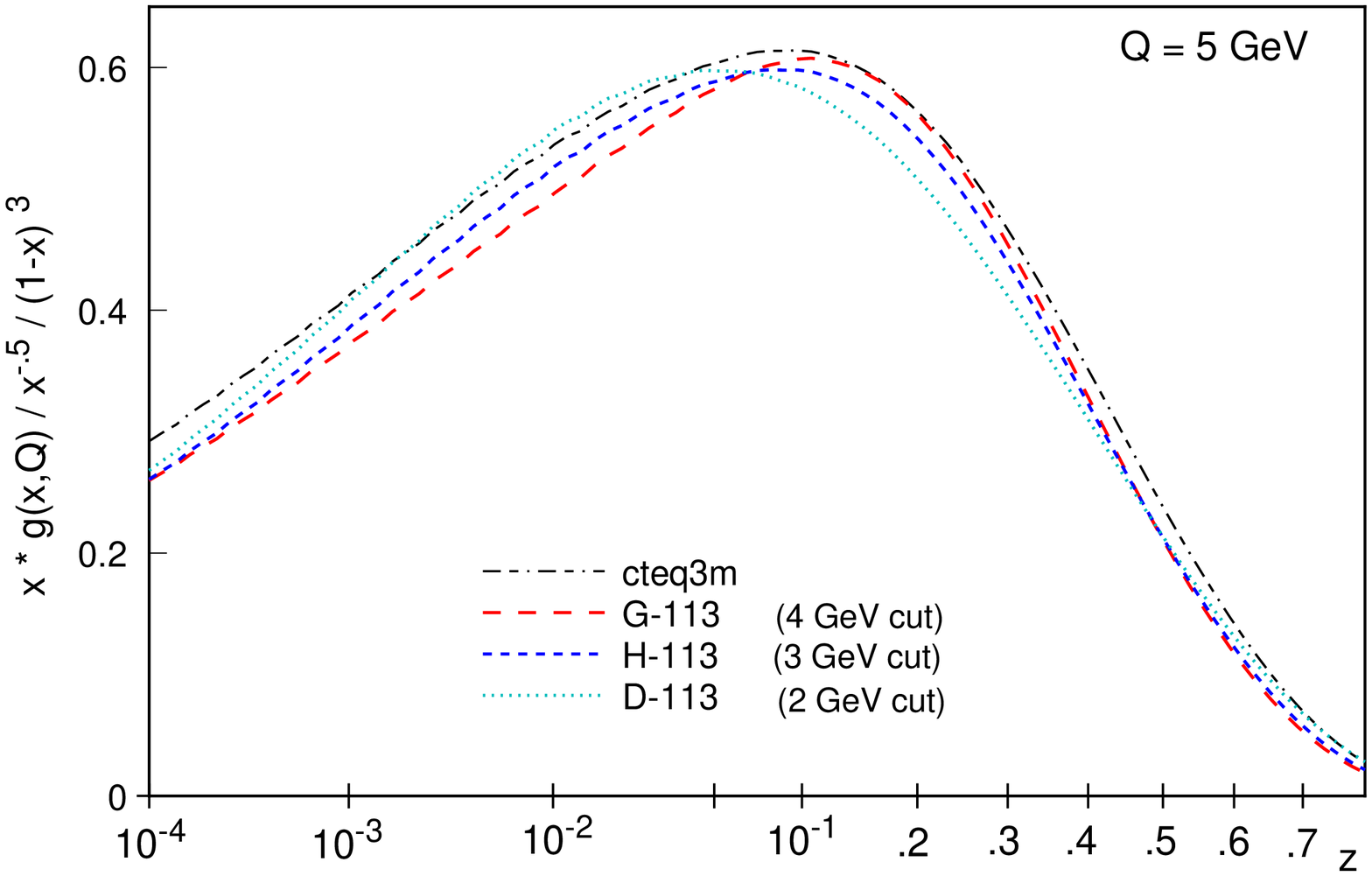}}

  \capt{: Comparison of gluon distributions obtained in three global fits
using three different values of $Q_{cut}$ in data selection.}
  \label{fig:QcutGlu}
\end{Figure}
}

\def\figJetHJ
{
\begin{Figure}
\epsfysize=3.2in
\centerline{\epsfbox{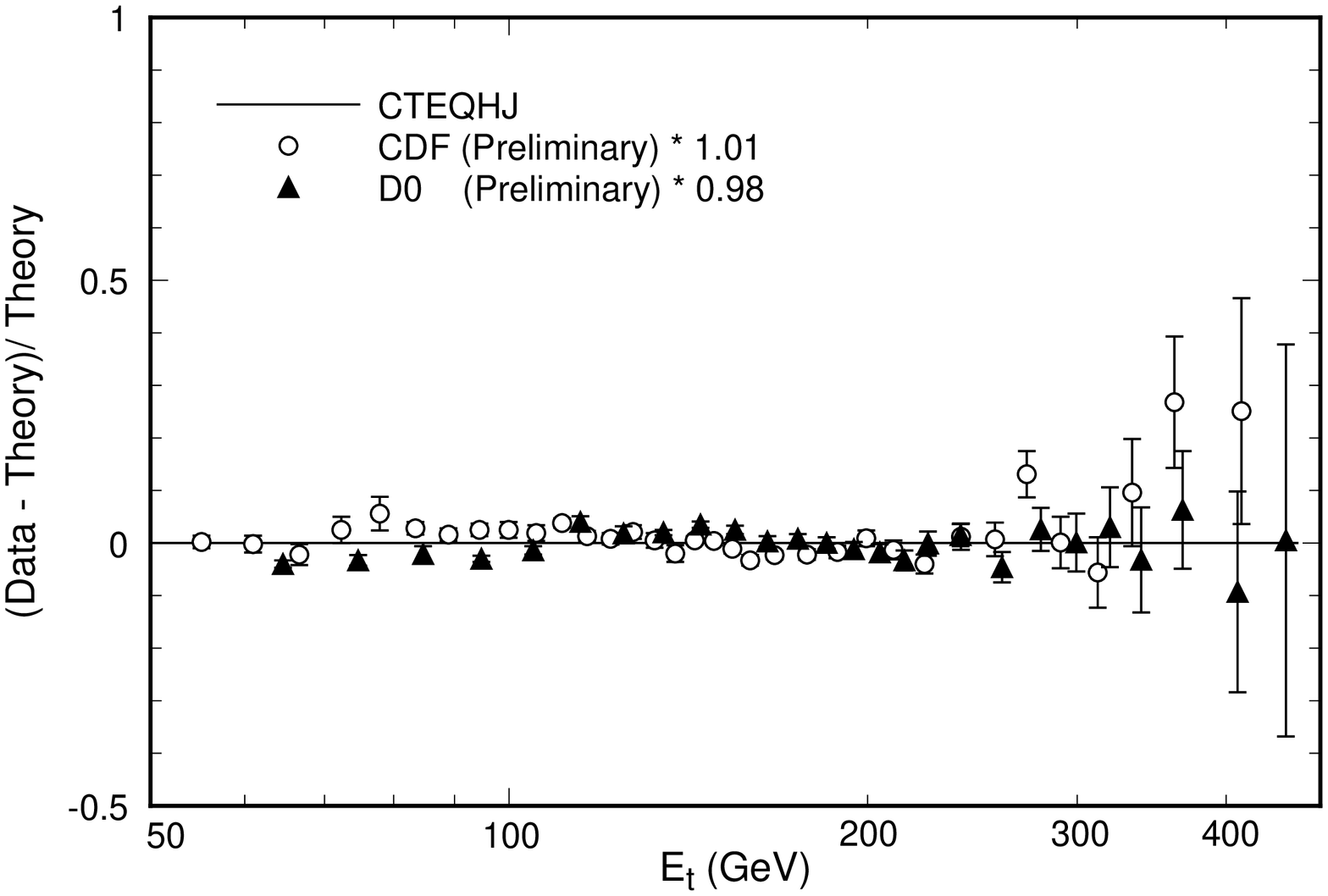}}

  \capt{: Inclusive jet cross-section of CDF and D0
compared to NLO QCD calculations based on the CTEQ4HJ parton
distributions.}
  \label{fig:JetHJ}
\end{Figure}
}

\def\figbcdms
{
\begin{Figure}
\epsfysize=4in
\centerline{\epsfbox{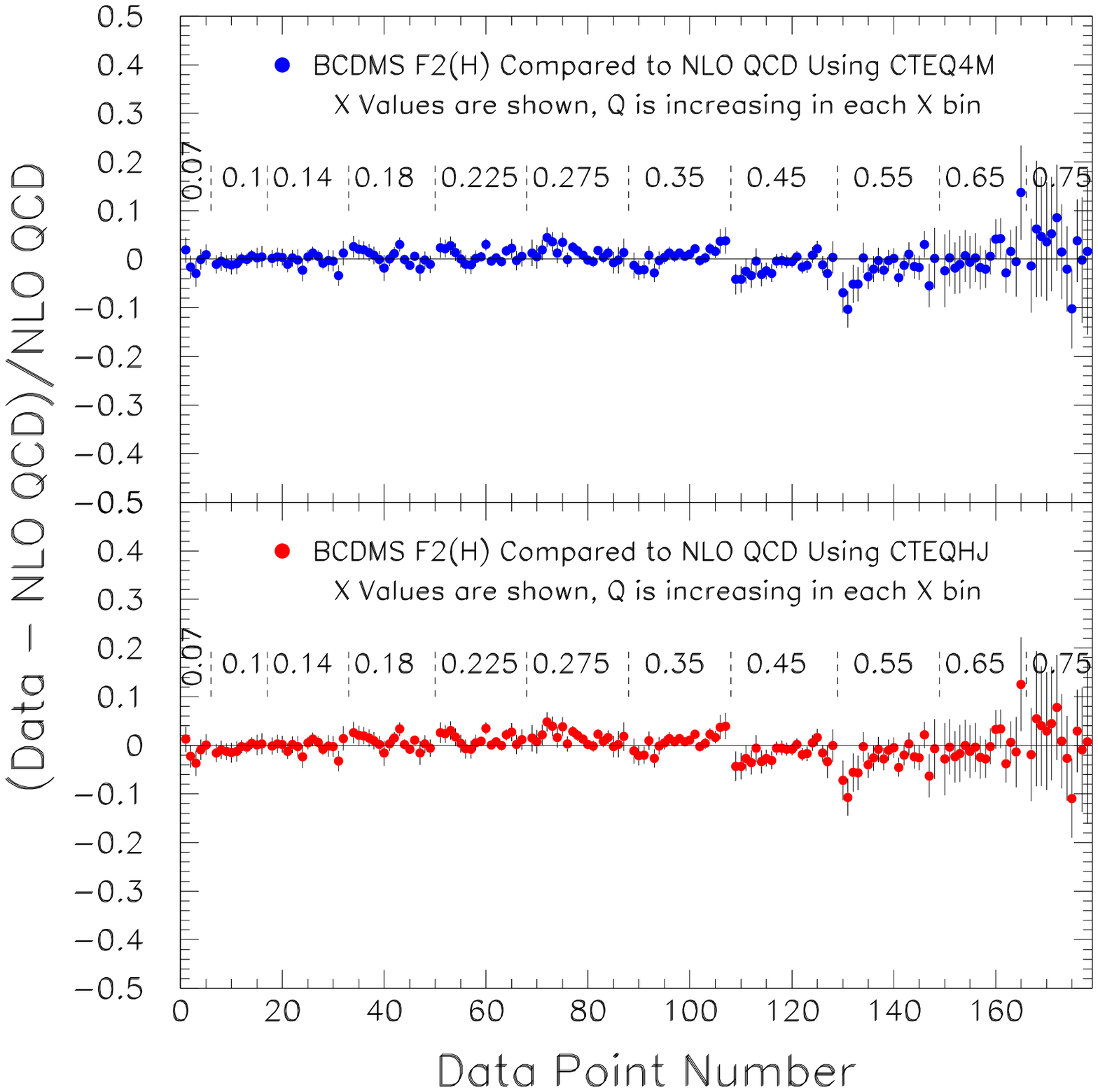}}

  \capt{: Percentage deviation of BCDMS proton data from NLO QCD values
based on CTEQ4M and CTEQ4HJ. Both PDF sets give good fits.}
  \label{fig:bcdms}
\end{Figure}
}

\def\figxGluSqk
{
\begin{Figure}
\epsfysize=3.2in
\centerline{\epsfbox{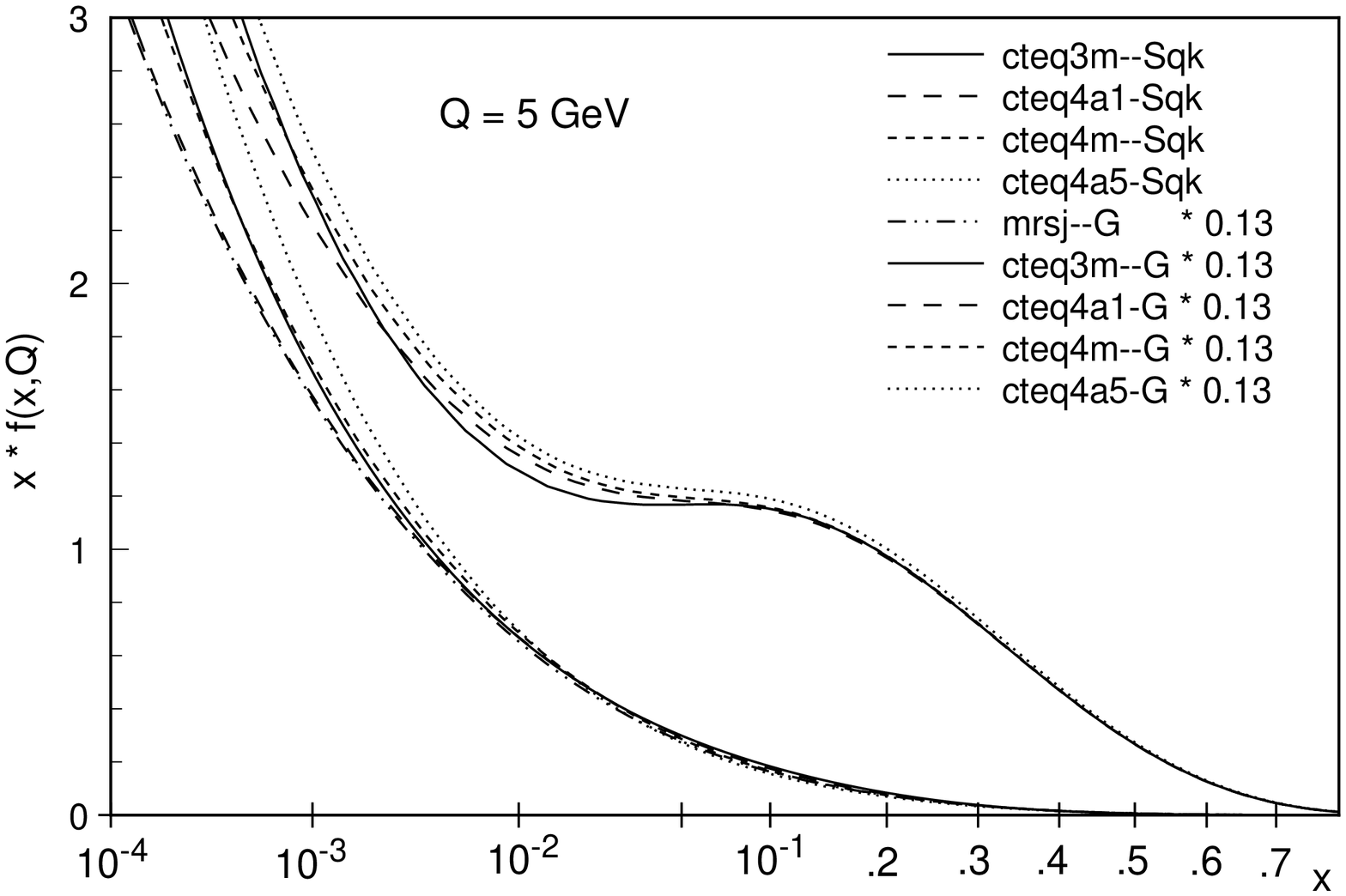}}

  \capt{: Comparison of $x\,G(x,Q)$ and $x\,S(x,Q)$ between some new
parton distribution sets and those from CTEQ3M. $S(x,Q)$ is the singlet
quark distribution (sum over all flavors). The CTEQ3M and CTEQ4M gluons
appear to lie on top of each other. The same is true for the CTEQ4A1 and
MRSJ gluons. Differences in $G(x,Q)$ for $x>0.01$ are not evident in
this plot.}
  \label{fig:xGluSqk}
\end{Figure}
}


\def\tblFits
{
\begin{table}[h]
\begin{center}
\begin{tabular}{|c|c|c|c|c|}
\hline
Series & New  & Inclusive & parame- & Section \\
& DIS data & Jet Data & trization & discussed \\ \hline\hline
A &  &  & m & \ref{sec:nojet} \\ \hline
B & x &  & m & \ref{sec:nojet},\ref{sec:Jets} \\ \hline
C & x &  & m+2 & \ref{sec:nojet} \\ \hline
CTEQ4A & x & x & m+2 & \ref{sec:CTEQ4} \\ \hline
Q$_{cut}$ & x & x & m & Appendix \\ \hline
\end{tabular}
\end{center}

\caption{Several series of global fits on which the physics discussions are
based. ``New DIS data'' refers to those becoming available since 1995.
Minimal parametrization ``m'' refers to Eq.~2; and ``m+2'' refers
to Eq.~3. The last column refers to the section number where the
specific series is discussed.}
\label{tbl:Fits}

\end{table}
}
\def\tblExptLis
{
\begin{table}
\begin{center}
\begin{tabular}{|c|c|c|c|c|}
\hline
Process & Experiment & Measurable & Data Points & Ref.
\\ \hline\hline
DIS & BCDMS & $F_{2\ H}^\mu, F_{2\ D}^\mu $ & 324 & \cite{bcdms} \\ \hline
& NMC & $F_{2\ H}^\mu, F_{2\ D}^\mu, F_{2\ n/p}^\mu $ & 297 & \cite{NewNmc} \\ \hline
& E665 & $F_{2\ H}^\mu, F_{2\ D}^\mu $ & 70 & \cite{E665} \\ \hline
& H1 & $F_{2\ H}^e $ & 172 & \cite{NewH1} \\ \hline
& ZEUS & $F_{2\ H}^e $ & 179 & \cite{NewZeus} \\ \hline
& CCFR & $F_{2\ Fe}^\nu, x\ F_{3\ Fe}^\nu $ & 126 & \cite{ccfr} \\ \hline
Drell-Yan & E605 & $sd\sigma /d\sqrt{\tau }dy$ & 119 & \cite{E605} \\ \hline
& NA-51 & $A_{DY}$ & 1 & \cite{NA51} \\ \hline
W-prod. & CDF & Lepton asym. & 9 & \cite{Wasym} \\ \hline\hline
Direct $\gamma $ & WA70 & $Ed^3\sigma /d^3p$ & 8 & \cite{wa70} \\ \hline
& UA6 & $Ed^3\sigma /d^3p$ & 16 & \cite{UA6} \\ \hline\hline
Incl. Jet & CDF & $d\sigma /dE_t$ & 36 & \cite{CdfJets} \\ \hline
& D0 & $d\sigma /dE_t$ & 26 & \cite{D0Jets} \\ \hline
\end{tabular}
\end{center}
  \caption{List of processes and experiments used in the Global analysis. 
  \label{tbl:ExptLis}}

\end{table}
}
\def\tblCTEQ4PDFs
{
\begin{table}
\begin{center}
\begin{tabular}{|c|c|c|c|}
\hline
PDF set & Description & $\alpha _s(m_z)$ & $Q_0^2\;($GeV$^2)$ \\ \hline\hline
&Standard Sets&&  \\ \hline
CTEQ4M & $\overline{MS}$ scheme & 0.116 & 2.56 \\ \hline
CTEQ4D & DIS scheme & 0.116 & 2.56 \\ \hline
CTEQ4L & Leading Order & 0.132 & 2.56 \\ \hline\hline
&$\alpha_s$ series&&  \\ \hline
CTEQ4A1 &  1 & 0.110 & 2.56 \\ \hline
CTEQ4A2 &  2 & 0.113 & 2.56 \\ \hline
CTEQ4A3 & Same as CTEQ4M & 0.116 & 2.56 \\ \hline
CTEQ4A4 &  4 & 0.119 & 2.56 \\ \hline
CTEQ4A5 &  5 & 0.122 & 2.56 \\ \hline\hline
&Specials&&  \\ \hline
CTEQ4HJ & ``Hi-Jet'' & 0.116 & 2.56 \\ \hline
CTEQ4LQ & ``Low Q$_0^{}$'' & 0.114 & 0.49 \\ \hline
\end{tabular}
  \caption{List of new CTEQ4 parton distributions and their characteristics.}
  \label{tbl:CTEQ4PDFs}
\end{center}
\end{table}
}
\def\tblChiSqA
{
\begin{table}
\begin{center}
\begin{tabular}{|c|c||c|c|c|c|c|}
\hline
 Expt.     &  \#pts &  CTEQ4M     &  CTEQ4HJ     &  CTEQ4LQ    &  MRSJ        \\
\hline\hline
$BCDMS^H$  &   168  & 144.8(0.86) & 173.0(1.03) & 139.4(0.83) & 183.1(1.09) \\ \hline
$BCDMS^D$  &   156  & 185.6(1.19) & 205.9(1.32) & 182.5(1.17) & 229.3(1.47) \\ \hline
$NMC^H$    &   104  &  97.3(0.94) &  91.7(0.88) &  96.0(0.92) & 113.4(1.09) \\ \hline
$NMC^D$    &   104  &  93.3(0.90) &  90.2(0.87) &  97.9(0.94) & 122.7(1.18) \\ \hline
$NMC_R$    &    89  & 130.8(1.47) & 133.5(1.50) & 132.6(1.49) & 142.4(1.60) \\ \hline
$E665^H$   &    35  &  41.3(1.18) &  38.5(1.10) &  44.5(1.27) &  37.8(1.08) \\ \hline
$E665^D$   &    35  &  32.3(0.92) &  33.5(0.96) &  34.3(0.98) &  29.8(0.85) \\ \hline
$CCFR F_2$ &    63  &  83.2(1.32) &  72.4(1.15) &  74.3(1.18) & 107.7(1.71) \\ \hline
$CCFR F_3$ &    63  &  46.5(0.74) &  45.5(0.72) &  49.9(0.79) &  57.8(0.92) \\ \hline
$ZEUS$     &   179  & 243.4(1.36) & 232.7(1.30) & 268.5(1.50) & 252.4(1.41) \\ \hline
$H1$       &   172  & 118.9(0.69) & 120.2(0.70) & 131.9(0.77) & 109.6(0.64) \\ \hline
$CDF A_W$  &     9  &   4.3(0.48) &   3.4(0.38) &   3.8(0.42) &   3.3(0.37) \\ \hline
$NA51$     &     1  &   0.6(0.63) &   0.5(0.49) &   0.4(0.41) &   2.5(2.47) \\ \hline
$E605$     &   119  &  97.7(0.82) & 101.6(0.85) & 100.4(0.84) &  97.8(0.82) \\ \hline
\hline
 Total     &  1297  &   1320      &   1343      &   1356      &   1490      \\ \hline
\end{tabular}
\end{center}
  \caption{Total $\chi^2$ values and their distribution among the DIS and DY
experiments for current generation of parton distributions which take into
account the most recent HERA (1996) and NMC (1995) data.
In parentheses are the $\chi^2/point$ values.}
  \label{tbl:ChiSqA}
\end{table}
}
\def\tblChiSqB
{
\begin{table}
\begin{center}
\begin{tabular}{|c|c||c|c|c|c|}
\hline
Expt. & \#pts & MRSA' & CTEQ3M & MRSA & GRV \\ \hline
\hline
 BCDMS$^H$  &  168 & 156.9(0.93)& 128.7(0.77) & 168.0(1.00) & 250.3(1.49) \\ \hline
 BCDMS$^D$  &  156 & 213.7(1.37)& 190.3(1.22) & 215.3(1.38) & 187.2(1.20) \\ \hline
 NMC$^H$    &  104 & 129.0(1.24)& 146.6(1.41) & 114.4(1.10) & 123.8(1.19) \\ \hline
 NMC$^D$    &  104 & 151.8(1.46)& 137.3(1.32) & 135.2(1.30) & 115.4(1.11) \\ \hline
 NMC$_R$    &   89 & 143.3(1.61)& 134.4(1.51) & 140.6(1.58) & 129.0(1.45) \\ \hline
 E665$^H$   &   35 &  38.2(1.09)&  47.6(1.36) &  37.8(1.08) &  39.9(1.14) \\ \hline
 E665$^D$   &   35 &  29.1(0.83)&  44.5(1.27) &  29.5(0.84) &  29.8(0.85) \\ \hline
 CCFR $F_2$ &   63 &  68.0(1.08)&  66.2(1.05) &  68.7(1.09) & 164.4(2.61) \\ \hline
 CCFR $F_3$ &   63 &  54.1(0.86)&  41.9(0.67) &  61.7(0.98) & 114.7(1.82) \\ \hline
 ZEUS       &  179 & 368.7(2.06)& 549.5(3.07) &1222.6(6.83) & 843.1(4.71) \\ \hline
 H1         &  172 & 149.5(0.87)& 220.2(1.28) & 407.6(2.37) & 404.2(2.35) \\ \hline
 CDF $A_W$  &    9 &   4.2(0.47)&   3.0(0.33) &   3.7(0.41) &   9.6(1.07) \\ \hline
 NA51       &    1 &   0.1(0.06)&   0.4(0.42) &  0.01(0.01) &   0.01(0.01) \\ \hline
 E605       &  119 &  93.5(0.79)&  92.6(0.78) &  95.9(0.81) &  90.3(0.76) \\ \hline	
 Total      & 1297 &   1600     &   1803      &   2701      &   2502      \\ \hline
\end{tabular}
\end{center}
  \caption{Total $\chi^2$ values and their distribution among the DIS and DY
experiments for the previous generation of parton distributions which includes
experimental data available in 1995 (MRSA') or before 1995 (CTEQ3M, MRSA). GRV
does not perform a full global fit. Since it is used widely, it is included here
for reference. In parantheses are the $\chi^2/point$ values.}
  \label{tbl:ChiSqB}
\end{table}
}
\def\tblMparam
{
\begin{table}
\begin{center}
\begin{tabular}{|c||c|c|c|c|c|c|}
\hline
 Parton & $A_0$ & $A_1$ & $A_2$ & $A_3$ & $A_4$ & \% Momentum \\ \hline \hline
$xd_v$  &  0.640 &  0.501 &  4.247 &  2.690 &  0.333 &   11.2\\ \hline
$xu_v$  &  1.344 &  0.501 &  3.689 &  6.402 &  0.873 &   30.6\\ \hline
$xg$  &  1.123 & -0.206 &  4.673 &  4.269 &  1.508 &   41.7\\ \hline
$x(\overline d - \overline u)$  &  0.071 &  0.501 &  8.041 &  0.000 & 30.000 & -- \\ \hline
$x(\overline d + \overline u)$  &  0.255 & -0.143 &  8.041 &  6.112 &  1.000 & 13.2\\ \hline
$xs$  &  0.064 & -0.143 &  8.041 &  6.112 &  1.000 &    3.3\\ \hline
\end{tabular}
\end{center}
  \caption{Parameters for the CTEQ4M initial parton distributions at
$Q_0=1.6$ GeV. Also, $\alpha_s(m_z)=0.116$, corresponding to
$\Lambda_5 = 202$ MeV.}
  \label{tbl:Mparam}
\end{table}
}
\def\tblDparam
{
\begin{table}
\begin{center}
\begin{tabular}{|c||c|c|c|c|c|c|}
\hline
 Parton & $A_0$ & $A_1$ & $A_2$ & $A_3$ & $A_4$ & \% Momentum \\ \hline \hline
$xd_v$  &  0.724 &  0.490 &  3.839 &  1.688 &  0.338 &   11.3\\ \hline
$xu_v$  &  1.528 &  0.490 &  3.554 &  6.448 &  1.162 &   30.4\\ \hline
$xg$  &  2.141 & -0.058 &  7.554 & 36.405 &  2.223 &   43.7\\ \hline
$x(\overline d - \overline u )$  &  0.054 &  0.490 &  7.200 &  0.000 & 30.000 & -- \\ \hline
$x(\overline d + \overline u )$  &  0.154 & -0.227 &  7.200 &  6.949 &  1.000 & 11.7\\ \hline
$xs$  &  0.038 & -0.227 &  7.200 &  6.949 &  1.000 &    2.9\\ \hline
\end{tabular}
\end{center}
  \caption{Parameters for the CTEQ4D initial parton distributions at
$Q_0=1.6$ GeV.
Also, $\alpha_s(m_z)=0.116$, corresponding to NLO $\Lambda_5 = 202$ MeV.}
  \label{tbl:Dparam}
\end{table}
}
\def\tblLparam
{
\begin{table}
\begin{center}
\begin{tabular}{|c||c|c|c|c|c|c|}
\hline
 Parton & $A_0$ & $A_1$ & $A_2$ & $A_3$ & $A_4$ & \% Momentum \\ \hline \hline
$xd_v$  &  0.702 &  0.443 &  4.003 &  2.433 &  0.622 &   10.9\\ \hline
$xu_v$  &  1.226 &  0.443 &  3.465 &  7.589 &  1.146 &   30.1\\ \hline
$xg$  &  0.854 & -0.305 &  3.666 &  1.846 &  1.968 &   41.8\\ \hline
$x(\overline d - \overline u )$  &  0.050 &  0.443 &  6.877 &  0.000 & 30.000 & -- \\ \hline
$x(\overline d + \overline u )$  &  0.201 & -0.200 &  6.877 &  5.644 &  1.000 &13.8\\ \hline
$xs$  &  0.050 & -0.200 &  6.877 &  5.644 &  1.000 &    3.5\\ \hline
\end{tabular}
\end{center}
  \caption{Parameters for the CTEQ4L initial parton distributions at
$Q_0=1.6$ GeV. Also, LO $\Lambda_5 = 181$ MeV}
  \label{tbl:Lparam}
\end{table}
}
\def\tblLQparam
{
\begin{table}
\begin{center}
\begin{tabular}{|c||c|c|c|c|c|c|}
\hline
 Parton & $A_0$  & $A_1$  & $A_2$  & $A_3$  & $A_4$  & \% Momentum \\ \hline \hline
$xd_v$  &  0.852 &  0.573 &  4.060 &  4.852 &  0.693 &   14.7  \\ \hline
$xu_v$  &  1.315 &  0.573 &  3.281 & 10.614 &  1.034 &   40.4  \\ \hline
$xg$    & 39.873 &  1.889 &  5.389 &  0.618 &  0.474 &   31.2  \\ \hline
$x(\overline d - \overline u )$  &  0.093 &  0.573 &  7.293 &  0.000 & 30.000 &    - \\ \hline
$x(\overline d + \overline u )$  &  0.578 &  0.143 &  7.293 &  1.858 &  1.000 &   11.7\\ \hline
$xs$   &  0.096 &  0.143  &  7.293 &  1.858 &  1.000 &    1.9    \\ \hline
\end{tabular}
\end{center}
  \caption{Parameters for the CTEQ4LQ initial parton distributions at
$Q_0=0.7$ GeV. Also, NLO $\Lambda_5 = 174$ MeV}
  \label{tbl:LQparam}
\end{table}
}
\def\tblExpNorA
{
\begin{table}
\begin{center}
\begin{tabular}{|c||c|c|c|c|}
\hline
Expt.       &CTEQ4M  &CTEQ4HJ  &CTEQ4LQ & MRSJ  \\ \hline\hline
$BCDMS$     & 0.988  & 0.983  & 0.993  & 0.978 \\ \hline
$NMC$       & 1.016  & 1.015  & 1.022  & 1.018 \\ \hline
$NMC_R$     & 1.000  & 1.000  & 1.000  & 1.000 \\ \hline
$E665$      & 1.013  & 1.027  & 1.009  & 1.041 \\ \hline
$CCFR$      & 0.976  & 0.971  & 0.983  & 0.968 \\ \hline
$ZEUS$      & 1.004  & 0.999  & 1.001  & 1.018 \\ \hline
$H1$        & 0.993  & 0.978  & 0.987  & 0.994 \\ \hline
$CDF A_W$   & 1.000  & 1.000  & 1.000  & 1.000 \\ \hline
$NA51$      & 1.000  & 1.000  & 1.000  & 1.000 \\ \hline
$E605$      & 1.076  & 1.051  & 1.075  & 1.070 \\ \hline
\end{tabular}
\end{center}
  \caption{List of normalization factors for the experiments which
minimize the $\chi^2$'s given in the corresponding $\chi^2$ table.}
  \label{tbl:ExpNorA}
\end{table}
}
\def\tblExpNorB
{
\begin{table}
\begin{center}
\begin{tabular}{|c||c|c|c|c|}
\hline
Expt.      & MRSA' & CTEQ3M& MRSA  & GRV   \\ \hline\hline
$BCDMS$    & 0.977 & 0.988 & 0.977 & 0.957 \\ \hline
$NMC$      & 1.019 & 1.005 & 1.018 & 0.988 \\ \hline
$NMC_R$    & 1.000 & 1.000 & 1.000 & 1.000 \\ \hline
$E665$     & 1.045 & 0.997 & 1.040 & 0.992 \\ \hline
$CCFR$     & 0.968 & 0.976 & 0.968 & 0.949 \\ \hline
$ZEUS$     & 1.023 & 0.995 & 1.099 & 0.878 \\ \hline
$H1$       & 0.988 & 0.957 & 1.030 & 0.836 \\ \hline
$CDF A_W$  & 1.000 & 1.000 & 1.000 & 1.000 \\ \hline
$NA51$     & 1.000 & 1.000 & 1.000 & 1.000 \\ \hline
$E605$     & 1.025 & 1.097 & 1.008 & 1.012 \\ \hline
\end{tabular}
\end{center}
  \caption{List of normalization factors for the experiments which
minimize the $\chi^2$'s given in the corresponding $\chi^2$ table.}
  \label{tbl:ExpNorB}
\end{table}
}
%
\newpage
\figNmcHacM
\figDIScdM
\figGluAa
\figGluA
\figGluB
\figGluC
\figJetData
\figJetUncer
\figJetcM
\figSubProc
\figJetBs
\figKinMap
\figJetdM
\figJetCdA
\figGluCdA
\figJetSys
\figxGluSqk
\figDISLQ
\figJetHJ
\figbcdms
\figQcutGlu

%
%
\tblFits
\tblExptLis
\tblCTEQ4PDFs
\tblChiSqA
\tblChiSqB
\tblMparam
\tblDparam
\tblLparam
\tblLQparam
\tblExpNorA
\tblExpNorB


\begin{thebibliography}{99}
\bibitem{alphaRange}
Some related papers use a wider range of $\alpha _s$, extending up to 0.125 or 0.130.
Such large values are in strong disagreement with the highly precise BCDMS and CCFR
experiments. Specifically, in a typical global fit involving around 1000 data points,
the $\chi ^2$ value increases by about 200 above the best fit, mostly from these two
experiments.
\bibitem{No105}
The set with $\alpha _s=0.105$ is not included here and in
subsequent comparisons because it is strongly disfavored by the
new H1 and ZEUS data included in the fit to the B- and C-series, by the
inclusive jet data used in the CTEQ4 fits, and by the LEP measurements
of $\alpha_s$ -- even if the total $\chi^2$ of the $\alpha _s=0.105$ fit on
DIS and DY data sets by themselves is comparable to that of the
$\alpha _s=0.122$ set.
\end{thebibliography}
\end{document}